\documentclass[12pt]{article}
\usepackage{natbib}
\usepackage{verbatim}
\usepackage{graphicx}
\usepackage{psfrag}
\usepackage{bm}
\usepackage{lscape}
\usepackage{longtable}
\usepackage{pdflscape,array,booktabs}
\usepackage{array}
\usepackage{amsmath}
\usepackage{amssymb}
\usepackage{xcolor}
\usepackage{makeidx}         
\usepackage{graphicx}        
\usepackage[bottom]{footmisc}
\usepackage{natbib}		
\usepackage{multirow}
\usepackage{subfigure}
\usepackage{authblk}
\newtheorem{theorem}{Proposition}

\def\tu{\tilde{u}}
\def\baru{\bar{U}}

\def\ca{c^\ast}
\def\cA{{\cal A}}
\def\chia{\chi^\ast}
\def\tg{\tilde{g}}
\def\tga{\tilde{g}^\ast}

\def\bcdot{\bm{\cdot}}
\def\barrho{\bar{\rho}}
\def\barrhow{\bar{\rho}_w}
\def\barrhowp{\bar{\rho}_w'}

\def\tubzero{\hat{u}_{b0}^\ast}
\def\tvbzero{\hat{v}_{b0}^\ast}
\def\twbzero{\hat{w}_{b0}^\ast}
\def\tubone{\hat{u}_{b1}^\ast}
\def\tvbone{\hat{v}_{b1}^\ast}
\def\twbone{\hat{w}_{b1}^\ast}
\def\tvbtwo{\hat{v}_{b2}^\ast}
\def\tuizero{\hat{u}_{i0}^\ast}
\def\tvizero{\hat{v}_{i0}^\ast}
\def\twizero{\hat{w}_{i0}^\ast}
\def\tuione{\hat{u}_{i1}^\ast}
\def\tvione{\hat{v}_{i1}^\ast}
\def\twione{\hat{w}_{i1}^\ast}

\def\tpizero{\tilde{p}_{i0}^\ast}
\def\tpione{\tilde{p}_{i1}^\ast}

\def\bu{{\bf u}}
\def\Re{\mbox{Re}}
\def\Ma{\mbox{Ma}}
\def\Pr{\mbox{Pr}}
\def\Fr{\mbox{Fr}}

\def\mub{\mu_b}
\def\bI{{\bf I}}
\def\kx{k_x}
\def\kz{k_z}
\def\kxzero{k_{x0}}
\def\kxone{k_{x1}^\ast}
\def\kzzero{k_{z0}}
\def\kzone{k_{z1}^\ast}
\def\kxzerosq{k_{x0}^2}
\def\kzzerosq{k_{z0}^2}
\def\trho{\hat{\rho}}

\def\tT{\hat{T}}
\def\tp{\hat{p}}

\def\kaxsq{k_x^{\ast 2}}

\def\tT{\tilde{T}}

\def\Maasq{\mbox{Ma}^{\ast 2}}
\def\barT{\bar{T}}

\def\Masq{\mbox{Ma}^2}

\def\kxsq{k_x^2}
\def\kzsq{k_z^2}
\def\barrho{\bar{\rho}}
\def\baru{\bar{u}}

\def\barT{\bar{T}}
\def\barp{\bar{p}}
\def\trho{\hat{\rho}}
\def\tu{\hat{u}}
\def\tw{\hat{w}}
\def\tv{\hat{v}}
\def\tva{\hat{v}^\ast}
\def\tuzero{\hat{u}_0}

\def\tvzero{\hat{v}_0}
\def\twzero{\hat{w}_0}

\def\tvzeroadj{\hat{v}_0^+}
\def\tvone{\hat{v}_1}
\def\tvtwo{\hat{v}_2}

\def\chizero{\chi_0}

\def\cone{c_1^\ast}
\def\conei{c_{1I}^\ast}
\def\conei{\mbox{Im}(\cone)}

\def\czero{c_0^\ast}
\def\tp{\hat{p}}

\def\tpizero{\hat{p}_{i0}}
\def\tpione{\hat{p}_{i1}}

\def\tT{\hat{T}}
\def\Re{\mbox{Re}}
\def\Ma{\mbox{Ma}}
\def\Pr{\mbox{Pr}}

\def\mub{\mu_b}
\def\muT{\mu_T}

\def\tbu{\hat{{\bf u}}}

\def\bu{{\bf u}}

\def\func#1{F\left(#1\right)}

\def\cL{{\cal L}}
\def\cLf{{\cal L}_f}
\def\cLs{{\cal L}_s}
\def\cLadj{{\cal L}^+}
\def\cLfadj{{\cal L}_f^+}
\def\cLsadj{{\cal L}_s^+}
\def\tuw{\tu_w^\ast}
\def\tvw{\tv_w^\ast}
\def\tww{\tw_w^\ast}
\def\tuwa{\tu_w^\ast}
\def\tvwa{\tv_w^\ast}
\def\twwa{\tw_w^\ast}
\def\tpw{\tp_w}
\def\tpwa{\tp_w^\ast}
\def\yw{y_w^\ast}
\def\ywsq{y_w^{\ast 2}}
\def\ywcu{y_w^{\ast 3}}
\def\yi{y_i^\ast}
\def\yisq{y_i^{\ast 2}}

\def\tTw{\tT_w}
\def\chizero{\chi_0}

\def\dgammaw{\bar{u}_w'}
\def\ddgammaw{\bar{u}_w''}

\def\tTw{\hat{T}_w^\ast}
\def\trhow{\hat{\rho}_w^\ast}
\def\tstar{\hat{\star}}

\def\cR{c_R}
\def\cI{c_I}

\def\cIcf{{\cal C}_f}
\def\cIxf{{\cal X}_f}
\def\cIzf{{\cal Z}_f}
\def\cIcs{{\cal C}_s}
\def\cIks{{\cal K}_s}

\begin{document}
\title{Stability of a high Mach number flow in a channel.}


\author[1]{M. Deka}
\author[1]{G. Tomar}
\author[2]{V. Kumaran}
\affil[1]{Department of Mechanical Engineering, Indian Institute of Science, Bangalore 560 012, India}
\affil[2]{Department of Chemical Engineering, Indian Institute of Science, Bangalore 560 012, India}
\date{}                     
\setcounter{Maxaffil}{0}
\renewcommand\Affilfont{\itshape\small}

\maketitle
\begin{abstract}
Modal instabilities in a flow through a channel at high Reynolds and Mach numbers are studied for three-dimensional perturbations. In addition to the Tollmien-Schlichting modes, there exist higher modes in a channel flow that do not have a counterpart in the incompressible limit. The stability characteristics of these higher modes, obtained through numerical calculations, are compared with boundary layer and Couette flows that have been previously studied. The dominant higher mode instabilities in a channel flow are shown to be viscous in nature, in contrast to compressible boundary layer modes. For general compressible bounded-domain flows, a necessary condition for the existence of neutral modes in the inviscid limit is obtained. This criterion is used to construct a procedure to determine a critical value of Mach number below which the higher modes remain stable. This criterion also delineates a range of angles of inclination of the wave number with respect to the flow direction which could go unstable at a specified Mach number. Asymptotic analysis is carried out for the lower and upper branch of the stability curve in the limit of high Reynolds number. A common set of relations are identified for these exponents for the upper and lower branch for the continuation of the Tollmien-Schlichting modes and the compressible modes. The scalings for the Tollmien-Schlichting modes are identical to those for an incompressible flow. The scalings for the finite wave number modes are different; the wave speed \( c \) scales as \( \Re^{-\frac{1}{3}} \) for the lower branch and \( \Re^{-\frac{1}{5}} \) for the upper branch, where \( \Re \) is the Reynolds number. The asymptotic analysis shows that the stability boundaries for three-dimensional perturbations at high Reynolds numbers can be calculated from the strain rate and the temperature of the base flow at the wall.
\end{abstract}

\section{Introduction}
\label{sec:introduction}

In compressible flow, the local thermodynamic state of the system influences the flow field and vice-versa. The flow field is determined by solving of the mass, momentum and energy equations simultaneously. To study the stability of such flows to small amplitude perturbations, the linearised equations for the perturbations have to be solved. Compared to an incompressible flow, the inclusion of the thermodynamic variables increases the dimension of the solution space. For perturbations in a steady plane-parallel flow, the additional solution space is spanned by normal modes which do not have a counterpart in the incompressible limit. These modes, referred to as ``compressible" or ``higher" modes, have been shown to determine the stability characteristics of several compressible shear flows at finite Mach numbers \citep[see][]{ref-lees_lin-46,ref-mack-84}.

The compressible boundary layer has been one of the most extensively studied shear flows \citep[see][]{ref-lees_lin-46,ref-mack-63,ref-mack-65,ref-mack-65b,ref-mack-84,ref-mack-87}. \cite{ref-mack-63,ref-mack-65,ref-mack-65b,ref-mack-84,ref-mack-87} demonstrated the existence of an infinite sequence of unstable modes in the inviscid limit when there is a relative supersonic region in the flow. The first of these modes, labelled as the $S_u$ family, show a monotonic increase in the real part of the wave  speed with the streamwise wave-number. The second, third and subsequent higher modes belong to the $S_{dn}$ family and show a monotonic decrease in the real part of the wave  speed with increasing streamwise wave-number. Unlike the incompressible boundary layer which is unstable due to a viscous instability, the first, second and higher modes of the compressible boundary layer are all unstable in the inviscid limit when the wall is adiabatic. For an isothermal (cooled) wall, the $S_u$ family (first mode) is stable in the inviscid limit, but the $S_{dn}$ family is unstable \citep[see][]{ref-mack-87}. At finite Mach numbers ($\approx 4$ and higher), the second mode is observed to be the most dominant instability; this result has been experimentally verified \citep[see][]{ref-kendall-75}. Similar higher mode instabilities have been reported for unbounded flows like compressible shear layers \citep{ref-blumen-1-70,ref-blumen-2-75,ref-shivamoggi-79,ref-mack-90}, mixing layers \citep{ref-greenough-89,ref-ragab-89,ref-tam-2-89} and plane jets \citep{ref-tam-1-89,ref-michalke-84,ref-kennedy-98}. While the higher modes for unbounded and semi-bounded compressible flows have similarities in their mathematical as well as physical description, the higher modes in bounded flows are qualitatively different.

The fundamental differences between the stability of bounded and unbounded compressible flows are consequences of the boundary conditions. For boundary layer flows, \cite{ref-mack-84} argued that the modal perturbations for compressible flows only require to be bounded in the free-stream, but need not decay to zero, unlike its incompressible counterpart. This can be demonstrated by solving the compressible Rayleigh equation (see appendix \ref{sec:inviscidflow} for derivation) in the free-stream for a zero pressure gradient boundary layer, where the base velocity and temperature are constants. The perturbation for the velocity component perpendicular to the surface far from the boundary layer can be expressed as \citep[see][]{ref-lees_lin-46},
\begin{equation}
\label{eq:intro1}
 \hat{v}(y) \sim A_1 \exp\left( (1 - \Ma_{r,\infty}^2)^{1/2} (\kxsq + \kzsq)^{1/2} y \right) + A_2 \exp\left( - (1 - \Ma_{r,\infty}^2)^{1/2} (\kxsq + \kzsq)^{1/2} y \right) ,
\end{equation}
where, $A_1$ and $A_2$ are constants, $\kx$ and $\kz$ are the streamwise and spanwise wave-numbers and $\Ma_{r,\infty}$ is the relative Mach number in the free-stream. The relative Mach number, $\Ma_r$, is defined as,
\begin{equation}
  \Ma_r = \dfrac{(\baru - c)}{\sqrt{\barT}} \Ma \:,
\end{equation}
where, $\baru$ and $\barT$ denotes the base flow velocity and temperature (scaled by the free-stream values) respectively, and $c = c_R + \imath\: c_I$, denotes the complex wave  speed for temporal waves. From equation \ref{eq:intro1}, it can be seen that for neutral modes (i.e. $c_I = 0$) with $\Ma_{r,\infty} > 1$ (supersonic waves), the amplitude of the normal velocity perturbation is a sinusoidal function of $y$, and hence has a non-vanishing amplitude in the limit of $y \rightarrow \infty$. In contrast, for bounded flows, all perturbations are constrained to vanish at the physical boundaries irrespective of whether the flow is subsonic or supersonic. Therefore, the conventional classification of the higher modes based on the relative Mach number is not relevant for bounded flows. 

This difference also has consequences for the extension of the classical Rayleigh theorem for compressible flows. From equation \ref{eq:intro1}, it can be seen that for supersonic neutral modes, the freestream perturbations are effectively linear combinations of the incoming and outgoing waves, whose relative amplitudes are denoted by the constants $A_1$ and $A_2$ (see \cite{ref-schmid-98} for details). \cite{ref-lees_lin-46} showed that for such waves with a critical point in the flow domain, the energy generated (or consumed) due to the Reynolds stress discontinuity across the critical point is effectively carried out (or in) to the boundary layer by these incoming and outgoing waves. Therefore, a generalised inflection point (equivalent to the inflection point in incompressible flows) at the same location as the critical point is only required for subsonic ($\Ma_r <1$) neutral modes. The inflection point criterion for bounded domains as derived by \cite{ref-duck-94} for a Couette flow, does not distinguish between the subsonic or supersonic nature of the wave. 

\cite{ref-duck-94}, therefore, categorised the higher modes of a compressible Couette flow into two families, based on the dependence of the real part of the wave  speed $c_R$ on the wave-number $\kx$. For the even (or lower) family of modes (mode II, IV, VI \ldots), $c_R$ increases monotonically with $\kx$ while the odd (or upper) family of modes (mode I, III, V \ldots), $c_R$ decreases monotonically with $\kx$. The odd family is qualitatively similar in this aspect with the mode II, III and higher modes of a compressible boundary layer, labelled as the $S_{dn}$ family. Similarly the even-family is similar to the first mode (labelled as $S_u$ family) of a compressible boundary layer flow, which is the only mode reported there that shows a monotonic increase in $c_R$ with increasing $\kx$. Beyond this, no similarities between the higher modes of the two cases are observed. An important difference arises in the the long wave limit ($\kx \rightarrow 0$) of these higher modes. To illustrate this, the compressible Rayleigh equation for two-dimensional perturbations (see section \ref{sec:inviscidflow} for details) can be integrated twice in the limit $\kx \rightarrow 0$, to obtain,
\begin{equation}
\label{eq:intro2}
  \hat{v}(y) = (\baru(y) - c) \int_{0}^y \dfrac{\barT(y^\prime) - \Ma^2 (\baru(y^\prime) - c)^2}{(\baru(y^\prime) - c)^2} dy^\prime,
\end{equation}
where the physical boundary with zero normal velocity is assumed to be located at $y=0$. For boundary layer flows, the wave  speed $c$ can be obtained by imposing the zero normal velocity boundary condition in the limit of $y \rightarrow \infty $. Assuming the boundary layer thickness $\approx y_\delta$, the integral in equation \ref{eq:intro2} can be expressed as the sum of a contribution from the boundary layer region and from the freestream region where the integrand is a constant, i.e.,
\begin{equation}
\label{eq:intro4}
  \lim_{y \rightarrow \infty} \hat{v} =  (1 - c) \int_{0}^{y_\delta} \dfrac{\barT(y^\prime) - \Ma^2 (\baru(y^\prime) - c)^2}{(\baru(y^\prime) - c)^2} dy^\prime + \dfrac{1 - \Ma^2 (1 - c)^2}{(1 - c)} \lim_{y \rightarrow \infty} (y - y_\delta) .
\end{equation}
Since the integral in the right hand side of equation \ref{eq:intro4} is finite, and normal velocity is bounded for $y \rightarrow \infty$ if the second term on the right vanishes, which implies the wave  speed is $c = 1 \pm {1}/{\Ma}$. These represent the long-wave limit of $c$ for the first and second mode instabilities of a boundary layer flow respectively \citep[see][]{ref-mack-84}. When the freestream conditions are supersonic, i.e. $\Ma>1$, the wave  speed $c = 1 - {1}/{\Ma}> 0$, in the limit of $\kx \rightarrow 0$ for the first mode.  Since the wave-number increases monotonically with wave  speed for this mode \citep[see][]{ref-mack-87}, for any small positive wave-number $c_R > 1 - 1/\Ma$, implying it becomes a subsonic mode ($\Ma_r < 1$). From the generalised inflection point criterion proposed by \cite{ref-lees_lin-46}, the subsonic mode can only be neutral at a unique wave-number where the wave  speed is equal to the base velocity at the inflection point. Therefore, the first mode becomes unstable at small wave numbers, thereby making it a long-wave instability. For flows in bounded domains, the normal velocity is zero at a finite value of $y$, and so the wave speed is not constrained. Imposing the zero normal velocity condition at both boundaries, equation \ref{eq:intro2} can be numerically solved to obtain the two values of the wave  speed $c$ that represent the limiting values for first and second modes for $\kx \rightarrow 0$ \citep[see][]{ref-duck-94}. For a compressible Couette flow, the first mode of the family of modes that exhibit an increase in $\cR$ with wave-number is mode II. Numerical calculations by \cite{ref-duck-94} have shown that in the limit of $\kx \rightarrow 0$, this mode approaches a finite negative wave  speed. Therefore at small wave-numbers, mode II is neutral ($c_I =0$) in the inviscid limit, since a negative wave  speed cannot produce a critical point singularity in $y \in (0,1)$. However, since the wave  speed increases with wave-number, the mode becomes non-neutral when the wave  speed passes through zero at a finite wave-number. Therefore, the mode II instability of a Couette flow is a finite wavelength instability. In summary, the key differences between the stability of bounded and unbounded domain are,
\begin{itemize}
 \item [(a)] There is a zero velocity condition for the velocity perturbation at two boundaries for a bounded flow, in contrast to the requirement that the velocity perturbation be bounded when the cross-stream co-ordinate tends to infinity for an unbounded flow. 
 \item[(b)] Due to the above, the relative Mach number is not a relevant parameter in determining the behaviour of compressible modes.
 \item[(c)] The compressible modes that show a monotonic increase in the real part of the wave-speed with wave-number become unstable at low wave-numbers for boundary layer flows, whereas this instability always occurs at finite wave-numbers for bounded flows.
 \end{itemize}

In the present work, the stability of a compressible flow driven in a plane channel by a constant body acceleration is studied. While the Couette flow configuration has been studied for a compressible flow, the flow in a channel has been studied previously only to understand the extension of the incompressible Tollmien-Schlichting mode to finite Mach numbers \citep[see][]{ref-xi-2014,ref-xi-2017}. The equivalent problem in the incompressible limit is the flow driven by a pressure gradient. The discrete eigenspectrum for an incompressible channel flow has three distinct branches - $A$ ($c_R \rightarrow 0$), $P$ 
($c_R \rightarrow 1$), and $S$ ($c_R \approx 2/3$), with the $A$-branch containing one unstable eigenvalue, referred to as the 
Tollmien-Schlichting (TS) instability (see \cite{ref-drazin_reid-98,ref-schmid-98} for details). The Tollmien-Schlichting mode 
is a viscous instability, as the base flow for this case does not have an inflection point, which is a necessary criteria for the existence of inviscid instabilities \citep[see][]{ref-rayleigh-1880}. However, in the limit of $\Re \rightarrow \infty$, the two-dimensional TS 
mode approaches the solution of the inviscid equation in the limit of the wave-number $\kx \rightarrow 0$ and wave  speed $c 
\rightarrow 0$. The corresponding normal velocity eigenfunction $ \hat{v}(y) \rightarrow 1- y^2$, which is the base flow velocity 
profile \citep[see][]{ref-makinde-2003}. This has been observed in the numerical solutions of the Orr-Sommerfeld equations at high 
Reynolds numbers; this issue is discussed in some detail in this study. While the Orr-Sommerfeld equation cannot be solved analytically 
for a plane channel flow, matched asymptotic expansions have been used for obtaining approximate solutions in the limit of $\kx \Re \rightarrow \infty$ (see \cite{ref-heisenberg-24}, \cite{ref-tollmien-29,ref-tollmien-47} 
and \cite{ref-lin-45,ref-lin-45b,ref-lin-46,ref-lin-55,ref-lin-57a}). Also global expansions of WBKJ type 
have been adopted to obtain the approximate solutions for the Orr-Sommerfeld equations \citep{ref-heisenberg-24}. 
These solutions can then be used to obtain eigenvalue relations in reasonable agreement with numerically computed eigenvalues 
\citep[see][]{ref-drazin_reid-98}. 

In general, the eigenfunctions in the inviscid limit have two kinds of singularities, a discontinuity at the boundary because the inviscid solutions do not satisfy the zero tangential velocity boundary condition, and a logarithmic singularity at a critical cross-stream location where the flow velocity is equal to the wave  speed, and the coefficient of the highest derivative becomes zero. When the critical point is away from the boundaries, the viscous corrections to these singularities appear in the form of an exponential wall layer correction of thickness $(\kx\Re)^{-1/2}$ and an Airy function correction of thickness $(\kx\Re)^{-1/3}$ across the critical point \citep[see][]{ref-schmid-98}. When the critical point is close to the wall, the nature of the viscous corrections at the wall and the scaling 
exponents change significantly. \cite{ref-smith-part1-79,ref-smith-part2-79} derived the scaling exponents for the multi-deck 
structure of the Tollmien-Schlichting mode at high Reynolds numbers using local matched asymptotic expansions. All typical viscous 
instabilities at high Reynolds numbers are characterized by a finite range of wave-numbers over which the growth rates are positive 
(or $c_I>0$). The wave-numbers where the growth rate passes through zero are called the lower and upper branch points. For 
the incompressible channel flow, both these points asymptotically tend to zero in the limit of $\Re \rightarrow \infty$ for the TS mode. \cite{ref-smith-part1-79} derived the exponents for the lower branch point as $\kx \sim \Re^{-1/7}$ and $c \sim 
\Re^{-2/7}$. In this region, the critical point is close to the wall such that the boundary and critical layers merge 
into a single viscous layer. For the upper branch point, the layers are separate and the exponents are $\kx \sim \Re^{-1/11}$ and $c 
\sim \Re^{-2/11}$ \citep{ref-smith-part2-79}. Analytical solutions of the compressible viscous linear stability equations are more difficult to obtain than the incompressible counterpart due to the increased dimension of the problem. \cite{ref-shivamoggi-78} used WBKJ type expansions similar to \cite{ref-heisenberg-24} to obtain the solution of the compressible stability equations 
at high Reynolds numbers for a general base flow. Consistent multi-deck asymptotic matching procedures have been employed to determine the non-linear effects on the lower branch solution by \cite{ref-smith-non-linear-79b,ref-smith-non-linear-79a} and subsequently for the upper branch by \cite{ref-bodonyi-80,ref-bodonyi-82}. Multi-deck asymptotic analysis for non-linear development of neutral disturbances in compressible boundary layers have been shown in the works of \citep{ref-gajjar-90,ref-gajjar-94}, and the upper branch scalings in the work by \cite{ref-gajjar-89}. 

In this study, the temporal modal instabilities in a plane channel flow are examined. There are two categories of instabilities for a compressible channel flow, the continuation of the Tollmien-Schlichting instability to finite Mach numbers, referred to here as mode `0', and the `compressible' or `higher' modes that do not have a counterpart in the incompressible limit. Both of these modes are studied here, and it is found that the compressible modes are more unstable at high Mach numbers. In this study, it will be shown that the instabilities in a channel flow are qualitatively different from both a boundary layer and a Couette flow in a bounded channel. The formulation is described in Section \ref{sec:conservationlaws}, and a criterion for the existence of neutral modes in bounded domain flows is derived in the inviscid limit. Numerical studies to demonstrate the instabilities in a channel flow are presented in section \ref{sec:channelflow}. The instability characteristics of the compressible higher modes, which determine the stability at high Mach numbers, are also explained in section \ref{sec:channelflow}. Following this, in section \ref{sec:criticalmach}, the bounds on the wave-speed for non-neutral modes derived in appendix \ref{sec:inviscidflow} and the criterion for the existence of a neutral mode derived in section \ref{sec:conservationlaws} are used to obtain a critical Mach number below which the compressible higher modes in a bounded flow are always stable. Asymptotic analyses for the lower and upper branches of the instabilities in the channel flow are derived in section \ref{sec:asymptoticanalysis}. The unstable modes are classified on the basis of two criteria, lower/upper branch and small/finite wave number modes. The small wave number mode is the continuation of the Tollmien-Schlichting instability for an incompressible flow, and the finite wave number instabilities refer to the compressible modes. Relations between the scaling exponents for the wave number and growth rate are obtained for each combination. Using an adjoint method, the leading order solution for the wave  speed is obtained using asymptotic analysis. In contrast to earlier studies on incompressible flows, the present calculations are carried out for three-dimensional disturbances. Numerical computations based on the complete Navier-Stokes equations are used to verify the asymptotic results. Finally important conclusions are presented in section \ref{sec:conclusion}.

\section{Formulation of stability analysis}
\label{sec:conservationlaws}

For compressible flow of an ideal gas, the non-dimensionalized equations for mass, momentum and energy conservation are,
\begin{eqnarray}
 \frac{\partial \rho}{\partial t} + \nabla \bcdot (\rho \bu) = 0, \label{eq:eq21}
\end{eqnarray}
\begin{eqnarray}
 \rho \left( \frac{\partial \bu}{\partial t} + \bu \bcdot \nabla \bu \right) & = & - \frac{\nabla p}{\gamma \Ma^2} + \frac{1}{\Re} \nabla \bcdot ( \mu (\nabla \bu +  (\nabla \bu)^T)) \nonumber \\ & & \mbox{}
 + \nabla ((\mub - \tfrac{2}{3}) \mu \nabla \bcdot \bu) + \frac{1}{\Fr} \rho {\bf a}, \label{eq:eq22}
\end{eqnarray}
\begin{eqnarray}
 \rho \left( \frac{\partial T}{\partial t} + \bu \bcdot \nabla T \right)
& = & -(\gamma - 1) p \nabla \bcdot \bu + \frac{\gamma}{\Re \Pr} \nabla \bcdot (
\kappa \nabla T) \nonumber \\ & & \mbox{} + \frac{\gamma (\gamma-1) \Ma^2}{\Re}
(\mu (\nabla \bu + (\nabla \bu)^T \nonumber \\ & & \mbox{} + (\mub - \tfrac{2}{3}) (\nabla \bcdot \bu) \bI)) 
: (\nabla \bu), \label{eq:eq24} \\ & & \label{eq:eq23}
p = \rho T.
\end{eqnarray}
Here, Mach, Reynolds, Prandtl and Froude numbers are defined as,
\begin{equation}
 \Ma = \dfrac{u_0^\ast}{\sqrt{\gamma R T_0^\ast}}\:, \:\: \Re = \dfrac{\rho_0^\ast u_0^\ast h}{\mu_0^\ast}\:, \:\: \Pr = \dfrac{\gamma R \mu_0^\ast}{(\gamma - 1) \kappa_0^\ast}\:, \:\: \Fr = \dfrac{u_0^{\ast 2}}{a^\ast h}\:,
\end{equation}
where $\rho_0^\ast$, $u_0^\ast$, $T_0^\ast$, $\mu_0^\ast$, $\kappa_0^\ast$ and $a^\ast$ are the scales used to non-dimensionalise the density, velocity, temperature, viscosity, thermal conductivity and the body acceleration respectively. These scales are chosen based on the problem. To study the stability of plane shear flows, the density, temperature, pressure and velocity fields are expressed as normal modes imposed on a base flow,
\begin{eqnarray}
 \star & = & \bar{\star}(y) + \star'(x, y, z) \nonumber \\
 & = & \bar{\star}(y) + \tstar(y) \exp{(\imath \kx (x - c t) + \imath \kz z)}
 \label{eq:fouriermodes}
\end{eqnarray}
where \( \bar{\star}(y) \) is the base state value, \( \star'(x, y, z) \) is the small
perturbation imposed on the base state, \( \tstar(y) \) is the amplitude of the 
Fourier mode, \( \kx \) and \( \kz \) are the wave numbers
in the \( x \) (flow) and \( z \) (span-wise) directions respectively, and \( c \) is the 
wave speed. Perturbations
are stable/unstable for $c_I \lessgtr 0$.

\begin{figure}
  \begin{center}
  \subfigure[{$\bar{u}$ vs. $y$}]{\includegraphics[width=2.6in]{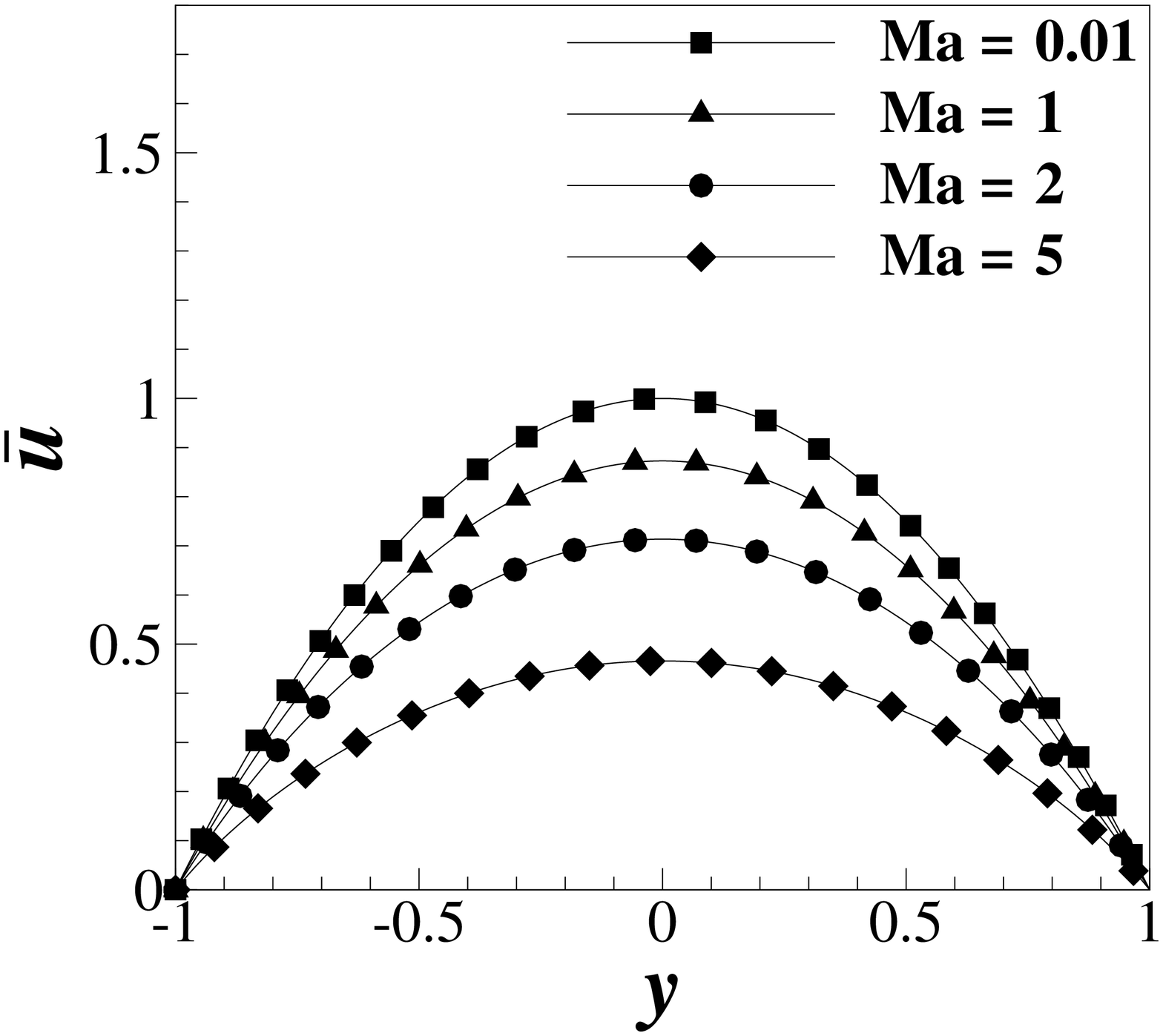}}
  \subfigure[{$\bar{T}$ vs. $y$}]{\includegraphics[width=2.6in]{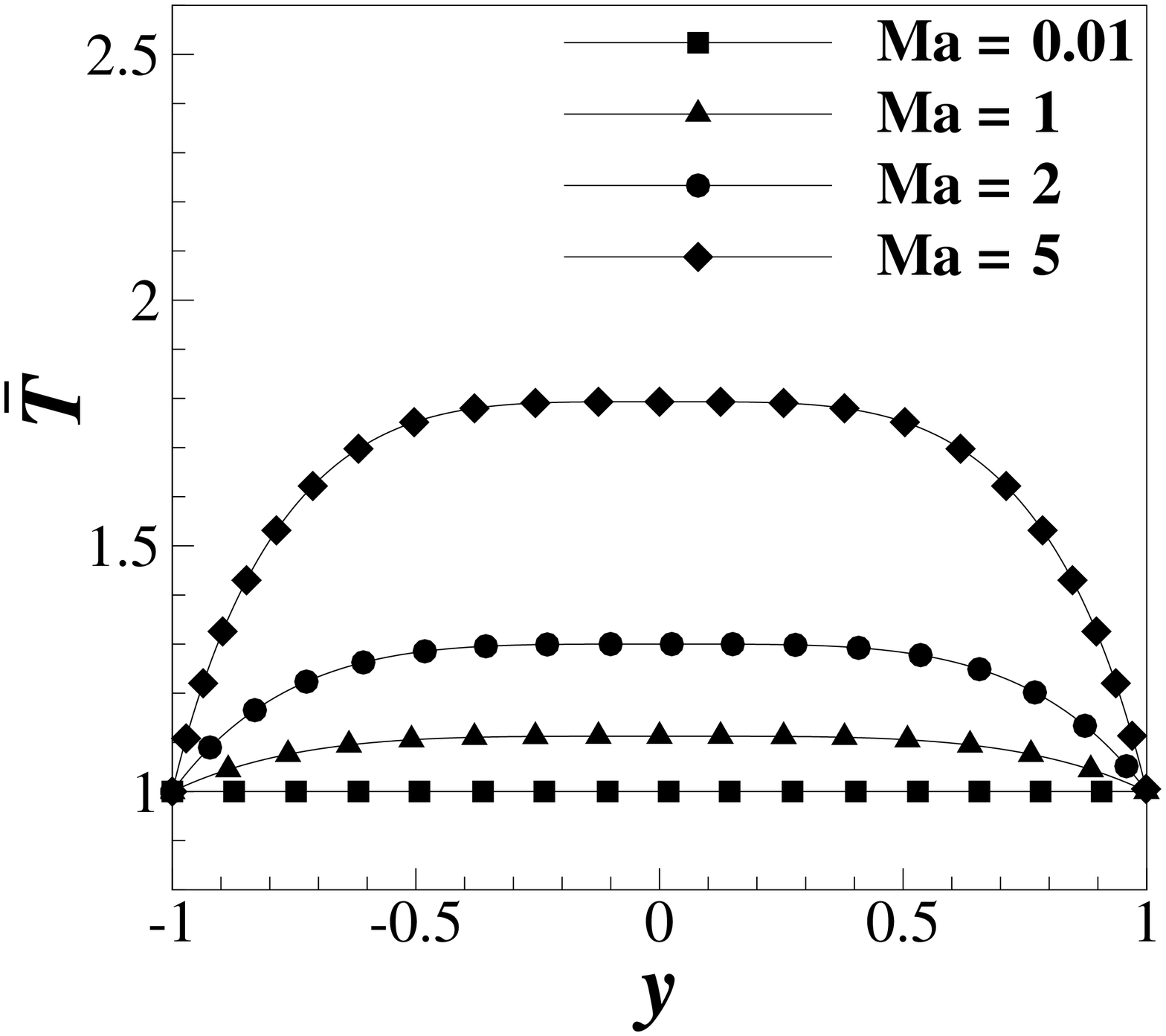}}
  \caption{{The velocity and temperature profiles for the base flow at finite Mach numbers.}}
  \label{fig-base_state}
  \end{center}
\end{figure}

A high Mach number flow of an ideal gas, in a channel of height $2 h$ and of infinite length and width is considered here, where the flow is in the $x$ direction, the velocity gradient is in the $y$ direction and the $z$ direction is perpendicular to the plane of the flow. The flow is driven by a constant body acceleration, ${\bf a}^\ast=(a^\ast,0,0)$, along the $x$ direction. 
The walls are considered to be maintained at a constant temperature $T_w^\ast$, where the superscript $^\ast$ is used for the dimensional quantities. The non-dimensionalisation scale for temperature is chosen as the wall temperature, i.e., $T_0^\ast = T_w^\ast$. The shear and bulk viscosity in the dimensional form, $\mu^*$ and $\mub^*$, and thermal conductivity $\kappa^*$, are related to the absolute temperature $T^*$ as, $\mu^* = \muT \func{T^*}$, $\mub^* = \mub \muT \func{T^*}$ and $\kappa^\ast = \kappa_T \func{T^*}$. The reference values for viscosity and thermal conductivity are considered based on the reference temperature, i.e. $\mu_0^\ast = \muT \func{T_w^*}$, and $\kappa_0^\ast = \kappa_T \func{T_w^*}$. To obtain the reference density, the equation of state for the base flow is used. For a steady fully-developed flow in a channel, the $y$ momentum equation reduces to a zero pressure gradient condition in the $y$ direction. Therefore, base flow pressure ($\barp^*$) becomes constant across the channel and the reference density is obtained as, $\rho_0^\ast = \barp^* / R T_w^*$. This base flow pressure ($\barp^*$) is used to non-dimensionalise pressure in the governing equations, thereby reducing the base flow equation of state to $\barrho \barT = 1$. 
A balance between the viscous and body
acceleration is used to obtain the non-dimensionalization scale for velocity. Setting $\Re = \Fr/2$, we obtain the velocity scale, $u_0^\ast = \rho_0^\ast a^\ast h^{2}/2 \muT \func{T_w^\ast}$. For the base unidirectional flow, the mass conservation equation is trivially satisfied, and the cross-stream pressure gradient is zero in the $y$ momentum conservation equation. The $x$ momentum and energy equations are expressed in dimensionless form,
\begin{eqnarray}
 \dfrac{1}{\Re} \frac{d}{dy} \left( \func{\barT} \frac{d \baru}{d y} \right) + \dfrac{2}{\Re}\frac{1}{\barT} = 0, 
 \label{eq:meanmom} \\
 \frac{\gamma}{\Re \Pr} \frac{d}{dy} \left( \func{\barT} \frac{d \barT}{d y} \right) + \frac{\gamma (\gamma-1) \Ma^2}{\Re} \func{\barT} \left( \frac{d \baru}{d y} \right)^2 & = & 0. \label{eq:meanene}
\end{eqnarray}
The base flow is obtained by solving equations \ref{eq:meanmom} and \ref{eq:meanene} numerically, with the boundary conditions, 
\begin{equation}
 \baru = 0\:, \:\: \barT = 1\:,
\end{equation}  
at both walls. Figure \ref{fig-base_state} shows the computed base flow profiles. For all the calculations for the channel flow, the viscosity and thermal conductivity are modelled for a monoatomic ideal gas using the hard-sphere model, $\func{T} = T^{\frac{1}{2}}$.

Substituting \ref{eq:fouriermodes} for the velocity, temperature and pressure into the Navier-Stokes equations \ref{eq:eq21}-\ref{eq:eq23} and subtracting the base flow equations, the linearised mass, $x$ momentum, $y$ momentum, $z$ momentum and temperature equations for normal modes can be obtained in the form,
\begin{equation}
 \mathcal L \mathbf{\hat{q}} = -\imath \kx c \mathbf{\hat{q}} 
\end{equation} 
where, $\mathbf{\hat{q}} = [ \trho \:\tu \:\tv \:\tw \:\tT ]^T$ is the solution vector. The equations are shown in the appendix \ref{secapp:normalmodeeqs}. The equations for the perturbations are solved subject to the following boundary conditions at the walls,
\begin{equation}
 \tu = \tv = \tw = \tT = 0.
\end{equation}

In the inviscid limit, the normal mode equations can be combined to obtain the compressible equivalent of the Rayleigh equation, 
 \begin{eqnarray}
  \frac{d}{d y} \left( \frac{(\baru - c)}{\chi} \frac{d \tv}{d y} - \frac{\tv}{\chi} \frac{d \baru}{d y} \right) & = & \frac{(\kx^2 + \kz^2) (\baru - c) \tv}{\barT}, \label{eq:eq218_a}
 \end{eqnarray}
where $\chi$ is,
\begin{eqnarray}
 \chi & = & \barT - \dfrac{\mbox{Ma}^2 \kx^2 (\baru - c)^2}{\kx^2+\kz^2} \label{eq:218_b}
\end{eqnarray}
The extension of the classical theorems for inviscid stability have been derived for bounded domains in appendix \ref{sec:inviscidflow}. 

One important result, which has no equivalent in the theorems for incompressible flows, can be obtained for neutral modes ($\cI = 0$) that are non-singular in the inviscid limit (i.e. no critical points in the flow domain). To obtain the above, the compressible Rayleigh equation \ref{eq:eq218_a} is written in terms of the function $\tg = \tv/(\baru - c)$, 
\begin{eqnarray}
 \frac{d}{d y} \left( \frac{(\baru - c)^2}{\chi} \frac{d \tg}{d y} \right) & = & \frac{(\kxsq + \kzsq) (\baru - c)^2 
 \tg}{\barT},
\end{eqnarray}
This equation is multiplied by the complex conjugate $\tga$ and integrated across the channel,
\begin{eqnarray}
 \left. \left( \frac{(\baru - c)^2 \tga}{\chi} \frac{d \tg}{d y} \right) \right|_{y_l}^{y_h} -
 \int_{y_l}^{y_h} dy \left( \frac{(\baru - c)^2}{\chi} \left| \frac{d \tg}{d y} \right|^2
 + \frac{(\kxsq + \kzsq) (\baru - c)^2 |\tg|^2}{\barT} \right) & = & 0. \nonumber \\ & & 
 \label{eq:eq15_a}
\end{eqnarray}
The first term on the left is zero at both boundaries due to the zero normal-velocity condition. For neutral modes with no critical point in $y \in (y_l,y_h)$ i.e., the wave  speed outside the range of the minimum and maximum of the base flow velocity, the terms in the above integral are regular. Hence, for non-singular neutral modes to exist, it is clear that $\chi$ has to be less than zero somewhere in the flow. This results in a restriction on the wave speed for neutral modes.
\begin{theorem}
For a neutrally stable mode with \( c_I = 0 \),
\begin{eqnarray}
 \chi & = & \barT - \dfrac{\mbox{Ma}^2 \kx^2 (\baru - c)^2}{\kx^2+\kz^2} \label{eq:proposition5}
\end{eqnarray}
has to be negative somewhere in the flow for non-zero wave numbers.
\label{theorem:proposition5}
\end{theorem}
It should be noted that proposition \ref{theorem:proposition5} indeed applies to the entire range of $c_R$. For neutral modes with $\cR$ outside the range of $(\mbox{Min}(\baru), \mbox{Max}(\baru))$, the $\tg$ is analytic in the flow domain and hence the terms inside the integral in equation \ref{eq:eq15_a} are regular. For neutral modes with  $\mbox{Min}(\baru)<\cR<\mbox{Max}(\baru)$, proposition \ref{theorem:proposition1}, shown in appendix \ref{sec:inviscidflow}, establishes that such modes can only exist if the flow has a generalised inflection point (GIP) and the wave-speed for the neutral mode is equal to the base velocity at the GIP. In that case, the second Frobenius solution obtained about the critical point is analytic (see \cite{ref-lees_lin-46} for the general forms of the local solutions about the critical point), and therefore the terms in the integral of equation \ref{eq:eq15_a} are also regular. Therefore, proposition \ref{theorem:proposition5} holds in general for all neutral modes for flows in a bounded domain. 

In Proposition \ref{theorem:proposition2} (Appendix \ref{sec:inviscidflow}), an
extension of the theorem for an incompressible flow shows that non-neutral modes
exist only if the real part of the wave speed is between the minimum and maximum of the flow
velocity, $\mbox{Min}(\baru) \leq c_R \leq \mbox{Max}(\baru)$. When there is a 
transition from neutral to non-neutral modes, the imaginary part of the wave speed 
transitions from zero to non-zero values. {Therefore in the limiting case, an instability could be obtained when $\cR = \mbox{Min}(\baru)$ or $\cR = \mbox{Max}(\baru)$, which provides a necessary condition for the existence of an instability as,}
\begin{eqnarray}
 \mbox{Min}\left(\barT - \dfrac{\mbox{Ma}^2 \kx^2 (\baru - \mbox{Min}(\baru))^2}{\kx^2+\kz^2} \right)
 & < & 0, \nonumber \\
 \hfill \mbox{or} \hfill & & \nonumber \\
 \mbox{Min}\left(\barT - \dfrac{\mbox{Ma}^2 \kx^2 (\mbox{Max}(\baru) - \baru)^2}{\kx^2+\kz^2} \right)
 & < & 0.
 \label{eq:macriterion}
\end{eqnarray} 
This criterion can be used to determine whether the velocity and temperature profiles, $\barT$ and $\baru$,
are potentially unstable, and this is used to identify the minimum Mach number for compressible
modes in section \ref{sec:criticalmach}.

Neutral modes with the wave-speed outside the interval $(\mbox{Min}(\baru), \mbox{Max}(\baru))$, are 
present in all compressible shear flows in the inviscid limit. Proposition \ref{theorem:proposition2}
stipulates that $c_I$ is zero for these modes, in the inviscid limit. There is a correction to $c_I$ due to 
dissipation in the viscous boundary layer which stabilizes these modes. Therefore, these modes are 
always stable in the high Reynolds number limit, and the modes that are potentially unstable satisfy
the criterion $\mbox{Min}(\baru) < c_R < \mbox{Max}(\baru)$. 
In the next section, the instabilities in a channel flow at finite Mach numbers will be shown through numerical solutions of the stability equations.

\section{Instabilities in a channel flow}
\label{sec:channelflow}
 The compressible channel flow is unstable to modal perturbations at high Reynolds number. The instabilities are of two types - the extension of the incompressible Tollmien-Schlichting (T-S) instability and the compressible higher mode instabilities. The continuation of the T-S mode at finite Mach numbers is a viscous instability. At all Mach numbers, there exists a unique critical Reynolds number, above which the T-S mode becomes unstable over a range of wave-numbers. The range of unstable wave-numbers, bounded by a lower and an upper branch in the wave number-Reynolds number plane, approaches zero for the T-S mode in the limit $\Re \rightarrow \infty$. In this limit, setting \( \kx = \kz = 0 \) in the inviscid equation \ref{eq:eq15_a} in section \ref{sec:conservationlaws}, it can be easily seen that \( \tg = (\tv/(\baru - c)) \) is a constant. The eigenfunction for the normal velocity perturbation is, therefore, \( \tv \propto (\baru - c) \) in the inviscid approximation. Since \( \baru \) is zero at the walls due to the no-slip condition, the no-penetration condition \( \tv = 0 \) is satisfied at the walls only if the wave speed \( c \) is zero in the inviscid limit. 
\begin{figure}
  \centerline{\includegraphics[width=2.6in]{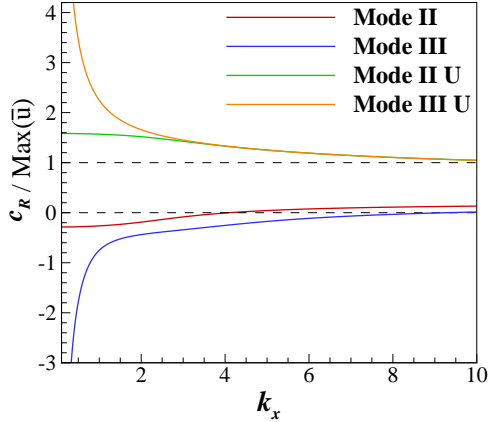}}
  \caption{The real part of the wave speed scaled with maximum velocity, as a function of wave number at \( \Ma = 2 \) for the two-dimensional inviscid modes 
  II and IV which are upstream traveling for \( \kx \rightarrow 0 \), and modes  I and III
  which are downstream traveling for \( \kx \rightarrow 0 \).}
 \label{fig-invisc-drp-ma-2}
 \end{figure}

The compressible modes are of two types, those that travel
downstream in the limit of low wave number and whose wave speed decreases as the wave number increases, and those that travel upstream in the limit of low wave number and whose wave speed increases as the wave number increases. The former, which
have been designated upper or odd modes and are numbered I, III, V, \ldots, while the latter are called lower or even 
modes and are numbered II, IV, VI, \ldots in \cite{ref-duck-94}, where a compressible Couette flow was studied. In the limit of low wave number, the wave speed of these 
modes is outside the range of the flow velocity, $c_R < \mbox{Min}(\baru)$ or $c_R > \mbox{Max}(\baru)$. 
From Proposition \ref{theorem:proposition2} in appendix \ref{sec:inviscidflow}, these modes are neutral, ($\cI=0$), since their wave  speed is not in the interval $(\mbox{Min}(\baru), \mbox{Max}(\baru))$. Therefore, these modes are neutrally
stable at low wave number in the inviscid limit, and these are stabilised by viscous effects at the wall.

Figure \ref{fig-invisc-drp-ma-2} shows the variation of the real part of the wave speed ($c_R$) with wave number ($k_x$) for two-dimensional inviscid modes ($k_z=0$) at \( \Ma = 2 \) for the channel flow. These are obtained by numerically integrating the compressible Rayleigh equation \ref{eq:eq218_a} across the channel by a procedure similar to \cite{ref-duck-94}. It can be seen in figure \ref{fig-invisc-drp-ma-2} that for $\kx \ll 1$, $c_R$ is outside the range $c_{R} < \mbox{Min}(\baru)$ for the lower modes, and $c_{R} > \mbox{Max}(\baru)$ for the upper modes. The wave speed tends to a finite value for the lowest modes in the limit $k_{x} \rightarrow 0$. This limiting value can be obtained by satisfying the zero normal velocity condition at both boundaries in the $\kx=0$ solution of the Rayleigh equation, given by equation \ref{eq:intro2} in section \ref{sec:introduction}. The magnitude of $\cR$ for the other modes diverge proportional to $\kx^{-1}$ in the limit $\kx \rightarrow 0$, similar to the modes in a Couette flow (\cite{ref-duck-94}).

\begin{figure}
  \begin{center}
    \subfigure{\includegraphics[width=2.6in]{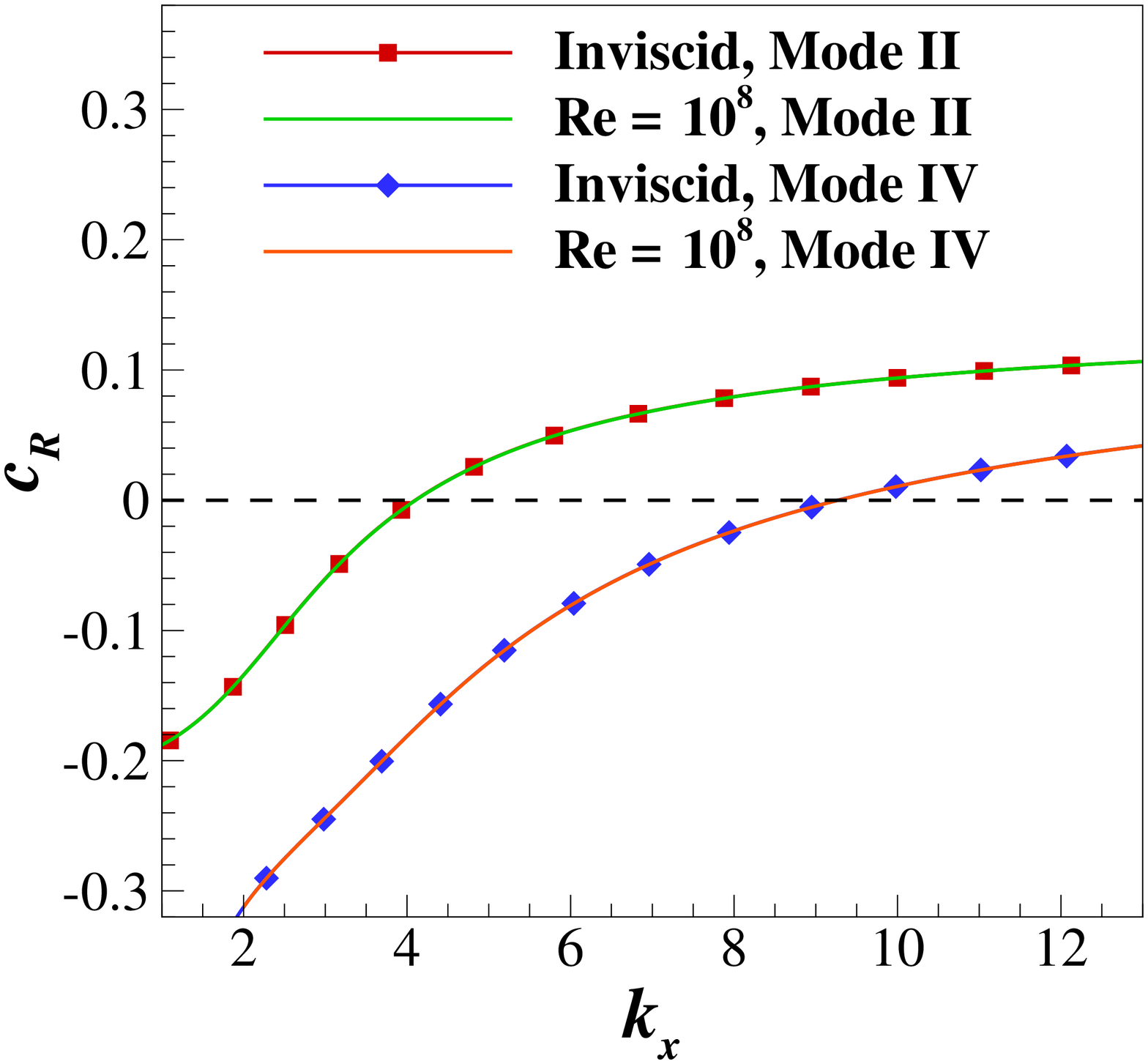}}
    \subfigure{\includegraphics[width=2.6in]{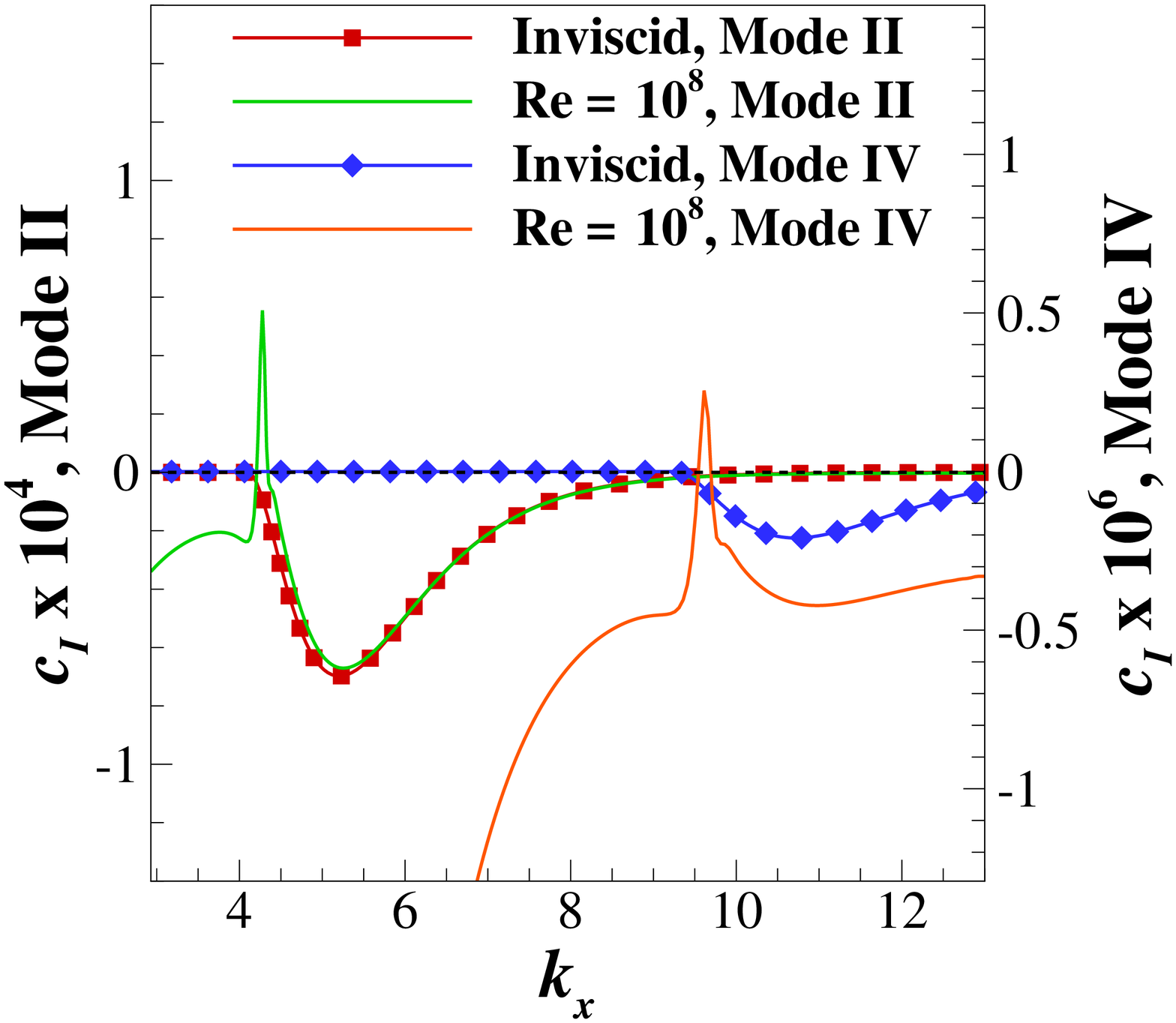}}
    \caption{The real part (a) and imaginary part (b) of the wave speed as a function of the stream-wise wave number for $k_z=0$ at Ma = 2 from the inviscid   analysis and from the complete linear stability equations at \( \Re = 10^8 \).} 
   \label{fig-invisc-drp-mode-2-3-comp}
   \end{center}
\end{figure}

As the wave number is increased, there exists a threshold wave-number $k_{x0}$ specific to each mode, above which the wave speed of the lower family of compressible modes increases above $\mbox{Min}(\baru)$, which is zero for the channel flow. This can be seen in figure
\ref{fig-invisc-drp-mode-2-3-comp} (a) where the variation of $c_{R}$ with $\kx$ is shown for the first two modes of the lower family, modes II and IV. Since the base state for the channel flow does not have a generalised inflection point, the higher modes in the inviscid limit are not neutral for all wave-numbers $\kx > k_{x0}$ (from proposition \ref{theorem:proposition1} in appendix \ref{sec:inviscidflow}). When $c_R$ for the lower mode is higher than $\mbox{Min}(\baru)$, proposition \ref{theorem:proposition2} does not apply, and the imaginary part of the growth rate $c_I$ should be non-zero. The inviscid calculation predicts
that $c_I$ is negative for a channel flow, as shown in figure \ref{fig-invisc-drp-mode-2-3-comp} (b). While figure \ref{fig-invisc-drp-mode-2-3-comp} (a) shows that the results of the inviscid calculation for $\cR$ are in quantitative agreement with those for the complete Navier-Stokes equations at $\Re = 10^8$, figure \ref{fig-invisc-drp-mode-2-3-comp} (b) clearly shows that the results for $\cI$ are not.
The inviscid calculation predicts that the lower modes are always stable, whereas it is observed to be unstable for a range of wave numbers
even at $\mbox{Re} = 10^8$, thereby indicating a viscous destabilization of the lower family modes.

While the variation of the wave speed with wave-number for the compressible modes of a channel flow are qualitatively similar to those of the Couette flow, there are important differences in the destabilisation mechanisms. \cite{ref-duck-94} has shown, by inviscid calculations for Couette flow, that when both walls are isothermal, both families of modes become stable ($\cI<0$) in the inviscid limit for $\kx > k_{x0}$. When the bottom wall is adiabatic and top is isothermal, the lower modes become unstable ($\cI>0$) while the upper modes become stable for $\kx > k_{x0}$. For a Couette flow, the minimum and maximum of the base velocity occur at the stationary and moving walls respectively. Therefore, in the viscous limit, for $\kx$ slightly greater than $k_{x0}$, the change in the nature of the wall-layer due to the emergence of a critical point close to the wall produces a viscous instability for both families of modes \citep[see][for the numerical results]{ref-hu-98}. For a channel flow, the minimum of base velocity (which is zero) occurs at both walls, whereas the maximum occurs in the interior of the domain. From the inviscid calculations, it has been observed that the upper family of modes are unstable ($c_I>0$) in the inviscid limit for $\kx > k_{x0}$. However, since the location of the maximum of the base flow velocity is within the domain, the upper (or odd) family of modes (unlike Couette flow) do not exhibit a viscous instability. This behavior of the odd modes for a channel flow are qualitatively similar to the $S_{dn}$ family of compressible boundary layer flows, as the maximum of the base velocity for a boundary layer flow is also located within the domain. The lower family of modes, as shown in the previous paragraph, are observed to be stable ($c_I <0$) from the inviscid calculations for $\kx > k_{x0}$. However, emergence of two critical points at both walls for $\kx > k_{x0}$ produce a viscous instability at high Reynolds number.

\begin{figure}
  \begin{center}
  \includegraphics[width=2.6in]{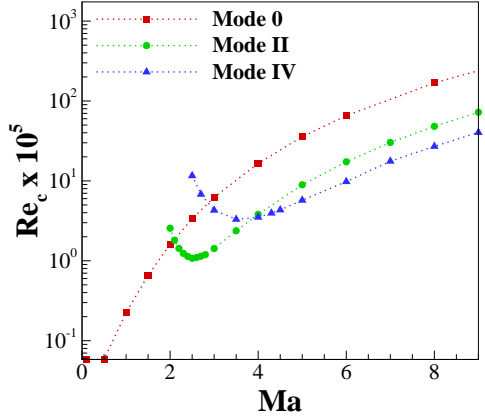}
  \caption{The critical Reynolds number for the two-dimensional modes 0, II and IV at different Mach numbers. }
  \label{fig-crit-val}
  \end{center}
\end{figure}

For all viscous instabilities, there exists a critical Reynolds number below which the damping effect of viscosity is sufficient to stabilise the mode at all wave-numbers. These critical Reynolds number are obtained from numerical calculations of the full viscous stability equations and shown in figure \ref{fig-crit-val} for modes 0 (finite Mach number extension of the T-S mode), II and IV. The critical Reynolds numbers for the upper family of modes have not been presented because being inviscid instabilities, it was observed from numerical calculations that for moderate Mach numbers ($\approx 2-9$), these modes were unstable at very high Reynolds numbers ($\gtrapprox 10^7$) only. Since, these values are at least an order of magnitude larger than the critical Reynolds numbers for the lower family viscous instabilities, the lower family determines the dominant instability characteristics at moderate Mach numbers.

From figure \ref{fig-crit-val} it can be seen that mode `0' is unstable at all Mach numbers but the critical Reynolds number increases with increasing Mach number indicating that compressibility has a stabilising effect on the finite Mach number extension of the T-S mode. This is consistent with the trends reported by \cite{ref-xi-2017}. The critical Reynolds numbers for the higher modes show a decrease first and then an increase with increasing Mach number. For a given mode, the Mach number corresponding to the critical Reynolds number is higher for higher modes. From figure \ref{fig-crit-val} it can also be seen that the critical Reynolds numbers increase rapidly with decreasing Mach numbers. The critical Reynolds numbers, in fact, approaches infinity at a specific finite value of Mach number, which indicates the existence of a critical Mach number below which the higher modes are always stable. In the next section, this critical Mach number for the compressible modes will be obtained using the theorems derived in the inviscid limit for the compressible flow in a bounded domain.

 \section{Critical Mach numbers}
 \label{sec:criticalmach}
 
 The numerical solution of the viscous stability equations indicate the existence of a finite Mach number at which the critical Reynolds number for the higher modes approach infinity, i.e., the higher modes are stable when Mach number is below this critical value. This bound on the Mach number can be obtained using proposition \ref{theorem:proposition5} shown in section \ref{sec:conservationlaws}. For higher modes, when $\cR$ is outside the range of the minimum and maximum of the base velocity, the mode is neutrally stable in the inviscid limit (see theorem \ref{theorem:proposition4} in appendix \ref{sec:inviscidflow}). These modes are stabilised when the effect of viscosity is included, because there is no critical point in the domain. This can be seen in the numerical solutions shown in figure \ref{fig-invisc-drp-mode-2-3-comp}. The instability, be it a viscous wall-layer instability (for the lower modes) or an inviscid instability (for the upper modes), can exist if $\mbox{Min}(\baru) \leq \cR \leq \mbox{Max}(\baru)$. The necessary condition for the existence of a neutral mode with $\cR$ outside the range $(\mbox{Min}(\baru),\mbox{Max}(\baru))$ in the inviscid limit is given by proposition \ref{theorem:proposition5}, which says that $\chi$ must be negative somewhere in the domain. In the limiting case, this implies that the minimum of $\chi$ must be less than or equal to zero for the existence of a neutral mode. This condition can be 
 re-written as,
 \begin{equation}
 \label{eq:crit_eq1}
   \mbox{Min}(\chi) = \barT_{0} -  \Ma^2 \cos^2 \psi (\baru_{0} - c)^2 \leq 0
 \end{equation}
where, the subscript $_0$ implies the value of the base flow variables at the point $y=y_0$, where $\chi$ is minimum, and the factor $\kxsq/(\kxsq + \kzsq)$ is written in terms of the wave-angle $\psi$, defined as,
 \begin{equation}
  \psi = \tan^{-1}\left(\dfrac{\kz}{\kx}\right)
 \end{equation}
 There are two possibilities for satisfying the condition \ref{eq:crit_eq1},
 \begin{equation}
 \label{eq:crit_eq3}
   c \leq \baru_0 - \dfrac{\sqrt{\barT_0}}{\Ma \cos \psi}\:,\:\: \mathrm{or}, \:\:c \geq \baru_0 + \dfrac{\sqrt{\barT_0}}{\Ma \cos \psi}.
 \end{equation}
 In the asymptotic limit of $\Ma \rightarrow 0$, the base flow temperature becomes a constant $\barT \rightarrow 1$, and the base flow velocity profile approaches the parabolic solution $\baru \rightarrow 1 - y^2$, which are both $O(1)$. Therefore, for $\cos{(\psi)} > 0$ in the limit of $\Ma \rightarrow 0$, 
 \begin{equation}
  c \sim \pm \dfrac{1}{\Ma},
 \end{equation}
 for equation \ref{eq:crit_eq3} to be satisfied. Therefore, as the flow approaches the incompressible limit, the wave  speeds for the lower and upper family modes approach negative and positive infinity respectively, and hence are not observed in the incompressible calculations. 
 
 As the Mach number increases from zero, the wave  speed for the lower (upper) family of modes increases (decreases) from negative (positive) infinity, as can be seen from the inequalities in equation \ref{eq:crit_eq3}. For the lower family of modes to remain neutral in the inviscid limit (and thereby stable in the viscous case), $c \leq \mbox{Min}(\baru)$. Therefore, in the limiting case, setting $c = \mbox{Min}(\baru)$ in equation \ref{eq:crit_eq1}, the critical value of Mach number for the lower family modes can be written as,
 \begin{equation}
 \label{eq:crit_eq5}
   \Ma_{cr,l} = \dfrac{\sqrt{\barT_0}}{\cos \psi (\baru_0 - \mbox{Min}(\baru))}
 \end{equation}
 Similarly, for the upper family of modes, the sufficient condition for stability is, $c \geq \mbox{Max}(\baru)$. Therefore, setting $c = \mbox{Max}(\baru)$ in equation \ref{eq:crit_eq1}, the critical Mach number for the upper family of modes is given by,
 \begin{equation}
 \label{eq:crit_eq6}
   \Ma_{cr,u} = \dfrac{\sqrt{\barT_0}}{\cos \psi (\mbox{Max}(\baru) - \baru_{0})}
 \end{equation}
 It is important to note that the location of the point $y_{0}$ in equations \ref{eq:crit_eq5} and \ref{eq:crit_eq6} are not the same. Since, $\chi$ is a function of $c$, the location of the point $y_0$, where $\chi$ is minimum, is dependent on the value of $c$ itself. Therefore, in order to calculate the critical Mach numbers from equations \ref{eq:crit_eq5} and \ref{eq:crit_eq6}, the location of the minima of $\chi$ has to be determined for $c = \mbox{Min}(\baru)$ and $c = \mbox{Max}(\baru)$, first. The critical Mach numbers can then be obtained from \ref{eq:crit_eq5} and \ref{eq:crit_eq6} through Newton-Raphson iterations, as the relations are implicit due to the dependence of the base flow on Mach number. 
  
To obtain the location of the minima for $\chi$, we can evaluate the slope of the function as,
\begin{equation}
\label{eq:chi_slope}
 \dfrac{d\chi}{dy} = \dfrac{d \barT }{dy} - 2 \Ma^2 (\baru - c) \dfrac{d\baru}{dy}
\end{equation} 
For a channel flow with isothermal walls, the base flow is symmetric, with $\mbox{Min}(\baru) = 0$ at the wall, and $\mbox{Max}(\baru)$ is the centerline velocity. In that case, it can be seen from equation \ref{eq:chi_slope} that $\chi$ increases monotonically from the wall to the centerline, for $c \geq \mbox{Max}(\baru)$. This implies that the $y_0$ to obtain the $\Ma_{cr,u}$ is at the wall. For the lower modes, it is not possible to show analytically, but numerical solutions of the base flow for all Mach numbers show that $y_0$ for $c \leq 0$ occurs at the center of the domain. Therefore, for the channel flow with isothermal walls, the critical Mach number expressions in equations \ref{eq:crit_eq5} and \ref{eq:crit_eq6} reduce to,
 \begin{equation}
 \label{eq:crit_eq7}
   \Ma_{cr,l} = \dfrac{1}{\cos \psi }\left(\dfrac{\sqrt{\barT}}{\baru}\right)\Bigg|_{y=0} \:,\:\: \Ma_{cr,u} = \dfrac{1}{\cos \psi }\left(\dfrac{1}{\mbox{Max}(\baru)}\right)
 \end{equation}
 For channel flow with one wall adiabatic, the symmetricity of the base flow is lost and the location of the minima of $\chi$ has to be determined numerically. Table \ref{tab:crit_ma} shows the critical Mach numbers calculated for the lower and upper family of modes for two-dimensional perturbations ($\psi = 0$), for a few possible bounded domain configurations. An interesting observation is that the critical Mach number for the lower family is higher when one wall is adiabatic.
 
\begin{table}
 \begin{center}
 \begin{tabular}{|l|l|l|l|}
 \hline
 Configuration            & Boundary conditions                                                                        & Lower modes & Upper modes \\ \hline
 \multirow{2}{*}{Channel} & both walls isothermal                                                                      & 1.329       & 1.192       \\ \cline{2-4} 
                          & \begin{tabular}[c]{@{}l@{}}one wall adiabatic\\ one wall isothermal\end{tabular}           & 7.601       & 2.164       \\ \hline
 \multirow{3}{*}{Couette} & both walls isothermal                                                                      & 1.00        & 1.00        \\ \cline{2-4} 
                          & \begin{tabular}[c]{@{}l@{}}moving wall isothermal\\ stationary wall adiabatic\end{tabular} & 1.00        & 1.134       \\ \cline{2-4} 
                          & \begin{tabular}[c]{@{}l@{}}moving wall adiabatic\\ stationary wall isothermal\end{tabular} & 1.134       & 1.00        \\ \hline
 \end{tabular}
 \caption{Critical Mach numbers for the higher modes for different bounded domain flows for two-dimensional perturbations ($\psi = 0$).}
 \label{tab:crit_ma}
 \end{center}
\end{table}
 
\begin{figure}
 \begin{center}
   \includegraphics[width=2.6in]{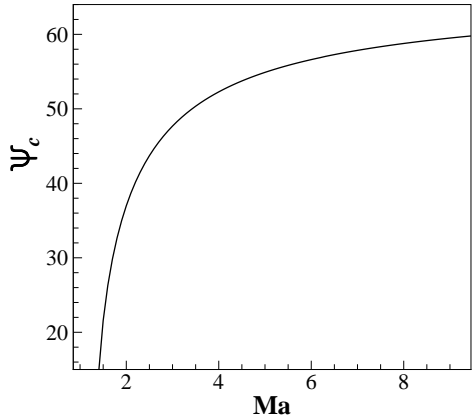}
   \caption{Critical wave-angle (in degrees) for the lower family modes in a channel flow with isothermal walls.}
  \label{fig-crit-psi}
  \end{center}
 \end{figure}

 From equation \ref{eq:crit_eq7}, it may be noted that the critical Mach numbers will be the lowest for two-dimensional modes, since $1/\cos\psi$ increases with increasing $\psi$. For three-dimensional modes at a specific wave-angle, a unique critical Mach number can be obtained. Conversely, one may obtain a critical wave-angle $\psi_c$ for a flow at a specific Mach number (above the critical Mach number for two-dimensional case), above which the flow will remain stable from equation \ref{eq:crit_eq7}. Figure \ref{fig-crit-psi} shows the critical wave-angle for various Mach numbers for the lower family modes in a channel flow with isothermal walls. The critical wave-angle increases with increase in Mach number.

\section{Asymptotic analysis}
\label{sec:asymptoticanalysis}

From the numerical results shown in section \ref{sec:channelflow}, the instabilities in a compressible channel flow are observed to be present at high Reynolds numbers. It was also shown that besides the extension of the T-S mode, the dominant instabilities at finite Mach numbers are those of the lower family of modes. All instabilities are characterised by two stability boundaries for each mode, the lower and the upper branch. The growth rate becomes positive, and the flow becomes unstable, when the lower boundary is crossed from the left in the wave-number axis.
When the upper boundary is crossed from the left, the growth rate becomes negative again (see figure \ref{fig-invisc-drp-mode-2-3-comp} (b)). Here, asymptotic analysis
is carried out for the lower and upper branches for mode `0' and the lower family compressible modes. The distinct structure of the eigenfunctions of these branches are examined, and the scalings of the viscous layer thickness and the viscous correction to the growth rate are derived. The non-dimensional velocity, pressure and wave
number are expressed as expansions in powers of \( \Re^{-1} \) in the asymptotic analysis. For clarity,
all terms are expressed in terms of scaled or `starred' variables with superscript \( ^\ast \)
multiplied by powers of \( \Re^{-1} \), where all starred variables are \( O(1) \) in the limit
\( \Re \gg 1 \).

\begin{figure}
 \begin{center}
  \includegraphics[width=0.5\textwidth]{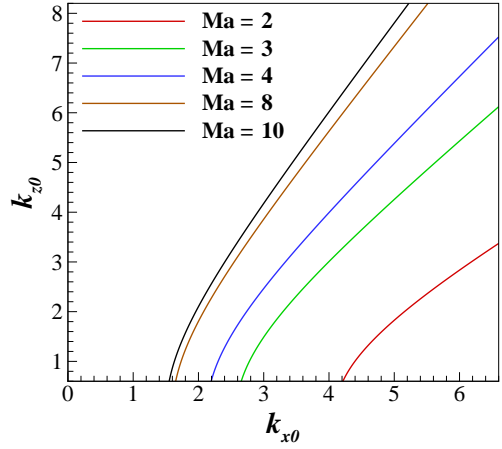}
  \caption{{$k_{x0}$ vs. $k_{z0}$ for inviscid standing wave solutions $( c = 0 )$ for mode II}}
 \label{fig-inv-neutral-3D}
 \end{center}
\end{figure}

There are two types of modes, those whose unstable wave numbers approach a finite value in the limit \( \Re \gg 1 \), called the finite wave number modes, and those for which the unstable wave numbers decrease to zero
in the limit \( \Re \gg 1 \), called the small wave number modes. The latter are
continuations of the Tollmien-Schlichting modes for an incompressible flow, while
the latter are present only in a compressible flow when one of the Mach number criteria, \ref{eq:macriterion}, is satisfied. The distinction is better understood
by examining equation \ref{eq:eq15_a}. For the low wave number continuation of the incompressible 
modes, the right side
is zero in the leading approximation in an expansion in the small parameter
$\Re^{-1}$. For these, equation \ref{eq:eq15_a} is satisfied because the wave number
is zero and $\tilde{g}$ is a constant, even though Proposition \ref{theorem:proposition5} is
not satisfied. In contrast, for the finite wave number modes, Proposition \ref{theorem:proposition5} 
and one of the Mach number criteria, \ref{eq:macriterion}, has to be satisfied. Due to this requirement,
these modes can become unstable only when the Mach number exceeds a critical value.
The expansions for the wave-number for the finite wave number `compressible' modes and the 
continuation of the incompressible modes are, respectively,
\begin{eqnarray}
 \kx & = & \kxzero + \Re^{- \alpha_k} \kxone, \: \: \: \kz = \kzzero + \Re^{- \alpha_k} \kzone,
 \label{eq:finitewavenumber} \\
 \kx & = & \Re^{- \alpha_k} \kxzero, \: \: \: \kz = \Re^{- \alpha_k} \kzzero.
 \label{eq:smallwavenumber}
\end{eqnarray}
For the finite wave-number modes, the wave-number pair ($k_{x0},k_{z0}$) corresponding to $c=0$ in the limit \( \Re \gg 1 \) are computed numerically by solving the Rayleigh equation and shown for different Mach numbers in figure \ref{fig-inv-neutral-3D}. The expansion for the wave speed is formulated as follows. There is a transition from stable to unstable
modes when the inviscid equation \ref{eq:eq218_a} becomes singular at the location where the wave speed 
\( c_R \) is equal to the flow velocity. It is necessary to incorporate viscous effects in a thin region
around this point in order to regularise the equation. In the high Reynolds number limit, the location
of the singularity approaches the wall, where \( \baru = 0 \), and the wave speed \( c_R \) is small
compared to the maximum velocity. Thus the expansion for the wave speed is,
\begin{eqnarray}
 c & = & \Re^{- \alpha_c} \czero + \Re^{- 2\alpha_c} \cone. \label{eq:wavespeed}
\end{eqnarray}

\subsection{Scaling relations}
\label{subsec:scalingrelations}
The scaling of the wave number in equation \ref{eq:smallwavenumber} and the wave speed in equation
\ref{eq:wavespeed} and the viscous wall layer thickness are determined from three considerations 
described in the following sub-sections. The modes are classified into finite wave number modes, for which equation 
\ref{eq:finitewavenumber} applies, and small wave number modes, for which equation \ref{eq:smallwavenumber}
applies. A distinction is also made between the lower branch and the upper branch of the 
stability curves. For the lower branch, there is a viscous layer at the wall within which the wave
speed is equal to the flow velocity. For the upper branch, flow velocity is equal to the wave speed
in an internal critical layer whose distance from the wall is small compared to the channel width, but
which is well separated from the viscous layer at the wall.
The relations between the exponents for the different types of modes is 
obtained in the following sub-sections. A summary of the relations between the different exponents is provided in 
table \ref{tab:tablescal}.
\subsubsection{Scaling from Rayleigh equation}
The stream-wise velocity in the bulk of the flow is considered to be \( O(1) \),
\begin{eqnarray}
 \tu & = & \tubzero + \Re^{- \alpha_u} \tubone. \label{eq:streamwise}
\end{eqnarray}
The magnitude of the cross-stream velocity in the bulk depends on the magnitude of the wave
number. For finite wave number modes, the stream-wise and cross-stream velocities are 
comparable, and the expansion for the cross-stream velocity is,
\begin{eqnarray}
 \tv & = & \tvbzero + \Re^{- \alpha_u} \tvbone. \label{eq:crossstreamfinite}
\end{eqnarray}
For finite wave number modes, the expansion for the Rayleigh equation \ref{eq:eq218_a} is,
\begin{eqnarray}
  \cLf(\tvbzero) & = & 0, \label{eq:rayleighfinitea} \\
  \Re^{-\alpha_u} \cLf(\tvbone) & = & \Re^{- \alpha_c} \czero \cIcf(\tvbzero) + \Re^{- \alpha_k}
  [\kxone \cIxf(\tvbzero) + \kzone \cIzf(\tvbzero)], \label{eq:rayleighfiniteb}
 \end{eqnarray}
 where the linear operator $\cLf$ (where the subscript \( _f \) denotes finite wave number modes) is defined as,
 \begin{eqnarray}
  \cLf(\tvbzero) & = & \frac{d}{d y} \left( \frac{\baru}{\chizero} \frac{d \tvbzero}{d y} - \frac{\tvbzero}{\chizero} \frac{d \baru}{d y} \right) - \frac{(\kxzero^2 + \kzzero^2) \baru \tvbzero}{\barT}. \label{eq:operatorfinite}
 \end{eqnarray}
 and the inhomogeneous terms in equation \ref{eq:rayleighfiniteb} are,
 \begin{eqnarray}
  \cIcf(\tvbzero) & = &  
  \frac{d}{dy} \left( \cA(\tvbzero)
  + \frac{1}{\chizero} \frac{d \tvbzero}{d y} \right)  - \frac{(\kxzero^2 + \kzzero^2) \tvbzero}{\barT}, \label{eq:inhomfinitec} \\
  \cIxf(\tvbzero) & = & \mbox{} 
  - \frac{d}{d y} \left( \frac{\kzzero \baru \cA(\tvbzero)}{\kxzero (\kxzero^2+\kzzero^2)}
  \right) 
  + \frac{2 \kxzero \baru \tvbzero}{\barT}, \label{eq:inhomfinitex} \\
  \cIzf(\tvbzero) & = & 
   \frac{d}{d y} \left( \frac{\baru \cA(\tvbzero)}{(\kxzero^2+\kzzero^2)}
   \right)
  + \frac{2 \kzzero \baru \tvbzero}{\barT},
  \label{eq:inhomfinitez}
 \end{eqnarray}
 where the function \( \cA \) is,
 \begin{eqnarray}
  \cA(\tvbzero) & = & \frac{2 \kxzero^2 \Masq \baru^3}{(\kxzero^2 + \kzzero^2) \chizero^2} \frac{d}{dy}
  \left( \frac{\tvbzero}{\baru} \right). \label{eq:inhomfinitea}
 \end{eqnarray}

 In deriving equations \ref{eq:inhomfinitec}-\ref{eq:inhomfinitez}, the function \( \chi \)
 is expanded in a series in the limit \( \Re \gg 1 \) as,
 \begin{eqnarray}
 \chi & = & \chizero + 2 \Masq \left(\frac{\Re^{- \alpha_c} \baru \czero \kxzero^2}{\kxzero^2+\kzzero^2} +
 \frac{\Re^{- \alpha_k} \baru^2 \kxzero \kzzero (\kxzero \kzone - \kxone \kzzero)}{(\kxzero^2 + \kzzero^2)^2}\right), \label{eq:chifinite}
\end{eqnarray}
where $\chizero = \barT - (\kxzero^2 \Ma^2 \baru^2 / (\kxzero^2 + \kzzero^2))$.

 If all terms in equation \ref{eq:rayleighfiniteb} are of same order, then we obtain the relation
 \( \alpha_c = \alpha_u = \alpha_k \) in the second row second column of table \ref{tab:tablescal}.
 \begin{table}
 \begin{center}
  \begin{tabular}{|l|c|c|} \hline
  Relation & Finite wave number & Small wave number \\ \hline
  First correction to & $\alpha_c = \alpha_u = \alpha_k$ &
  $\alpha_c = \alpha_u = 2 \alpha_k$ \\
  Rayleigh equation \ref{eq:eq218_a} & & \\ \hline
  Viscous layer stream-wise & $- \alpha_c = 2 \alpha_w - 1$ & $- \alpha_c - 
  \alpha_k = 2 \alpha_w - 1$ \\
  momentum equation \ref{u_mode} & & \\ \hline
   & Lower branch & Upper branch \\ \hline
   Normal velocity boundary & $\alpha_w = \alpha_u$ & $\alpha_w = 2 \alpha_u$ \\
   condition \( \tv = 0 \) at the wall & & \\ \hline
  \end{tabular}
 \caption{\label{tab:tablescal}The relations between the exponents 
 in the relations for the stream-wise velocity perturbations \( \tu = \tubzero + \Re^{- \alpha_u}  \tubone \), the wave number \( c = \Re^{- \alpha_c} \czero + \Re^{- 2 \alpha_c} \cone \), and the viscous
 wall layer thickness \( \delta_w = \Re^{- \alpha_w} \). The expansion for the wave number for 
 finite wave number modes is \( \kx = \kxzero + \Re^{- \alpha_k} \kxone \), \( \kz = \kzzero +
 \Re^{- \alpha_k} \kzone \). The expansion for the small wave number modes is \( \kx = \Re^{-
 \alpha_k} \kxzero, \kz = \Re^{- \alpha_k} \kzzero \).
 }
 \end{center}
 \end{table}
 
For small wave number modes, the cross-stream velocity scales as \( \kx \tu \), and the
expansion for the cross-stream velocity is,
\begin{eqnarray}
 \tv & = & \Re^{- \alpha_k} \tvbzero + \Re^{- \alpha_u - \alpha_k} \tvbone. \label{eq:crossstreamfinite}
\end{eqnarray}
The expansion for the Rayleigh equation, \ref{eq:eq218_a}, is
\begin{eqnarray}
  \Re^{- \alpha_k} \cLs(\tvbzero) & = & 0, \label{eq:rayleighsmalla} \\
 \Re^{- \alpha_u - \alpha_k} \cLs(\tvbone) & = & \Re^{- \alpha_c - \alpha_k} \czero \cIcs(\tvbzero) 
 \nonumber \\ & & \mbox{} +
 \Re^{- 3 \alpha_k} (\kxzero^2 + \kzzero^2) \cIks(\tvbzero)), 
 \label{eq:rayleighsmallb}
 \end{eqnarray}
 where the linear operator $\cLs$ (the subscript \( _s \) denotes small wave number modes) is defined as,
 \begin{eqnarray}
  \cLs(\tvbzero) & = & \frac{d}{d y} \left( \frac{\baru}{\chizero} \frac{d \tvzero}{d y} - \frac{\tvzero}{\chizero} \frac{d \baru}{d y} \right). \label{eq:operatorsmall}
 \end{eqnarray}
 and the inhomogeneous terms in equation \ref{eq:rayleighsmallb} are,
 \begin{eqnarray}
  \cIcs(\tvbzero) & = &  \frac{d}{dy} \left( \cA(\tvbzero)
+ \frac{1}{\chizero} \frac{d \tvbzero}{d y} \right), \label{eq:inhomsmallc} \\
  \cIks(\tvbzero) & = & \frac{\baru \tvbzero}{\barT}, \label{eq:inhomsmallk}
 \end{eqnarray}
 where \( \cA \) is defined in equation \ref{eq:inhomfinitea}.
 In deriving equations \ref{eq:inhomsmallc}-\ref{eq:inhomsmallk}, the expansion for \( \chi \)
 is,
 \begin{eqnarray}
 \chi & = & \chizero + 2 \Masq \left(\frac{\Re^{- \alpha_c} \baru \czero \kxzerosq}{\kxzero^2+\kzzerosq}
 \right), 
 \label{eq:chismall}
\end{eqnarray}
where \( \chizero = \barT - (\Masq \baru \kxzerosq/(\kxzerosq+\kzzerosq)) \).

 From equation \ref{eq:rayleighsmallb}, we obtain the relation \( \alpha_u = \alpha_c = 2
 \alpha_k \) in the second row third column of table \ref{tab:tablescal}.
The scaling relations are provided in the second row of table \ref{tab:tablescal}.

\subsubsection{Viscous wall-layer equations}
To satisfy the zero tangential velocity boundary condition, the stream-wise velocity near the wall requires viscous correction. In this work, we consider the total solution to be split into a bulk inviscid part and a viscous part that is non-zero only in the wall-layer region. 
For stream-wise velocity, the viscous wall-layer solution is designated \( \tuw \), and is considered to be \( O(1) \). If the viscous wall layer
thickness scales as \( \Re^{- \alpha_w} \), the cross-stream velocity is $\Re^{- \alpha_w}
\tvw$. In the stream-wise momentum conservation equation, the convective term is \( \barrho
\imath k (\baru - c) \tuw \sim \Re^{- \alpha_c} \), and the viscous term is \( \Re^{-1} (d^2
\tuw / d y^2) \sim \Re^{2 \alpha_w - 1} \). From a balance between these two terms, the 
relation \( - \alpha_c = 2 \alpha_w - 1 \), the third row second column in table 
\ref{tab:tablescal}, is obtained. 

For small wave number modes, the inertial term in the stream-wise momentum conservation
equation in the viscous sub-layer is \( \barrho \imath k (\baru - c) \tuw \sim \Re^{- \alpha_c
- \alpha_k} \). From this, we obtain the relation \( - \alpha_c - \alpha_k = 2 \alpha_w - 1 \), 
which is the third row third column in table \ref{tab:tablescal}.
The scaling relations from the viscous wall-layer are provided in the third row of table \ref{tab:tablescal}.
\subsubsection{No-penetration condition}
\label{subsubsec:nopenetration}
The boundary condition for the normal velocity perturbation \( \tv \) is satisfied by the 
solution \( \tvzero \) for the leading order Rayleigh equation, \ref{eq:rayleighfinitea} 
and \ref{eq:rayleighsmalla}. The leading order tangential velocity boundary condition
at the wall, \( \tubzero + \tuw = 0 \), provides a relation between the amplitude of the velocity
perturbations in the bulk and the wall layer. For the lower branch, the first correction for 
the normal velocity boundary condition is \( \Re^{- \alpha_u} \tvbone + \Re^{- \alpha_w} \tvw = 0 \).
The first correction to the velocity perturbation due to the bulk flow and the leading order
velocity due to the viscous wall layer are of equal magnitude for \( \alpha_u = \alpha_w \)
(fifth row second column of table \ref{tab:tablescal}) for the lower branch.

For the upper branch, the first correction to the normal velocity at the wall is also 
zero, \( \tvbone = 0 \). The second correction to the normal velocity boundary condition
is \(\Re^{- 2 \alpha_u} \tvbtwo + \Re^{- \alpha_w} \tvw = 0 \). Therefore, for the upper
branch, the relation is \( \alpha_w = 2 \alpha_u \) (fifth row third column of
table \ref{tab:tablescal}).

\subsection{Lower branch finite wave-number}
\label{subsec:lowerbranchfinite}

From table \ref{tab:tablescal}, the exponents for the lower branch for finite wave number
modes are \( \alpha_c = \alpha_u = \alpha_k = \alpha_w = \frac{1}{3}$. The wall layer thickness and the correction to the inviscid solution scale as \( \Re^{-\frac{1}{3}} \) for the lower branch. The scalings of the density, velocity, pressure and temperature fields are summarised in table \ref{tab:table1}. 
An interesting feature of the flow in the viscous wall layer is the scaling of the density,
temperature and pressure.
The pressure scaling in the viscous wall layer is determined from the stream-wise momentum conservation equation \ref{u_mode}, shown in appendix \ref{secapp:normalmodeeqs}, where \( \barrho \imath \kx (\baru - c)
\tu \sim \imath \kx \tp / (\gamma \Ma^2) \). If \( \gamma \) and \( \Ma \) are \( O(1) \), the 
pressure is \( O(\Re^{-\frac{1}{3}}) \). The temperature scaling in the wall layer is determined from the convective terms in the equation \ref{T_mode}, where \( \imath \kx (\baru - c) \tT \sim \tv (d 
\bar{T}/dy) \). From this, the temperature perturbation in the wall layer is \( O(1) \). From the 
linear approximation to the equation of state, \( \barrho \tT \sim \barT \trho \), the density
perturbation is also \( O(1) \). Thus, the pressure perturbation is \( O(\Re^{- \frac{1}{3}}) \)
smaller than the density and temperature perturbations, and the \( O(1) \) approximation for
the equation of state is \( \barrho \tT + \barT \trho = 0 \). The following analysis shows that 
it is not necessary to evaluate the density and temperature perturbations in order to determine
the stability, and so this is not pursued further.
\begin{table}
\begin{center}
 \begin{tabular}{|l|c|c|} \hline
  & Finite wave-number mode & Small wave-number mode \\ \hline
  & $\alpha_u = \alpha_c = \alpha_k = \alpha_w = \frac{1}{3}$ & $\alpha_u = \alpha_c = \alpha_w = 
  \frac{2}{7}, \alpha_k = \frac{1}{7}$ \\ \hline
  $\kx$ & $\kxzero + \Re^{-\frac{1}{3}} \kxone$ & $\Re^{-\frac{1}{7}} \kxzero$ \\ 
    $\kz$ & $\kzzero + \Re^{-\frac{1}{3}} \kzone$ & $\Re^{-\frac{1}{7}} \kzzero$ \\ 
    $c$ & $\Re^{-\frac{1}{3}} \czero$ & $\Re^{-\frac{2}{7}} \czero$ \\ \hline
    \multicolumn{3}{|c|}{Bulk flow} \\ \hline
$\tu$ & $\tubzero + \Re^{-\frac{1}{3}} \tubone$ & $\tubzero + \Re^{-\frac{2}{7}} \tubone$ \\ 
$\tv$ & $\tvbzero + \Re^{-\frac{1}{3}} \tvbone$ & $\Re^{-\frac{1}{7}} \tvbzero + 
\Re^{-\frac{3}{7}} \tvbone$ \\ 
$\tw$ & $\twbzero + \Re^{-\frac{1}{3}} \twbone$ & $\twbzero + \Re^{-\frac{2}{7}} \twbone$ \\ \hline
\multicolumn{3}{|c|}{Viscous wall layer} \\ \hline
$y$ & $\Re^{-\frac{1}{3}} \yw$ & $\Re^{-\frac{2}{7}} \yw$ \\ 
$\baru$ & $\Re^{-\frac{1}{3}} \baru_w' \yw$ & $\Re^{-\frac{2}{7}} \baru_w' \yw$ \\ 
$\tu$ & $\tuw$ & $\tuw$ \\ 
$\tv$ & $\Re^{-\frac{1}{3}} \tvw$ & $\Re^{-\frac{3}{7}} \tvw$ \\ 
$\tw$ & $\tww$ & $\tww$ \\ 
$\tp$ & $\Re^{-\frac{1}{3}} \tpw$ & $\Re^{-\frac{2}{7}} \tpw$ \\ 
$\tT$ & $\tTw$ & $\tTw$ \\ 
$\trho$ & $\trhow$ & $\trhow$ \\ \hline
 \end{tabular}
\caption{\label{tab:table1} Scalings for the cross-stream co-ordinate, velocity, density, temperature and pressure fields
for the lower branch asymptotic analysis.}
\end{center}
\end{table}

 Equation \ref{eq:rayleighfinitea} has to be solved numerically, with zero velocity conditions at the walls, to determine $\tvbzero$. The solution  close to the wall, which is necessary for imposing the boundary conditions for the normal and tangential velocity, is determined using an expansion of the 
 mean velocity and temperature close to the wall. Using the expansions \( \baru = \dgammaw y \) 
 close to the wall, the solution for \( \tvbzero \) is,
 \begin{eqnarray}
  \tvbzero & = & y, 
  \label{eq:eq47a} \\
  \kxzero \tubzero + \kzzero \twbzero & = & \imath. \label{eq:eq47b}
 \end{eqnarray}
 Here, the normalisation condition used is that the slope of \( \tvbzero \) is \( 1 \) at the wall. This imposes no loss of generality, since the amplitude of any one of the eigenfunctions can be set arbitrarily, and the amplitudes of the others are determined in terms of this one.

The velocity, density and pressure fields for the viscous wall layer from table \ref{tab:table1}
are substituted into the mass and momentum equations, \ref{rho_mode}-\ref{T_mode} (appendix \ref{secapp:normalmodeeqs}), and the 
largest terms in an expansion in \( \Re^{-\frac{1}{3}} \) are retained, to obtain the 
viscous wall layer equations. The co-ordinate in the wall layer, \( \yw = \Re^{-\frac{1}{3}} y\), 
is a stretched co-ordinate defined in table \ref{tab:table1}.
\begin{eqnarray}
  \imath \kxzero \tuwa + \frac{d \tvwa}{d \yw} + \imath \kzzero \twwa & = & 0, \label{eq:eq48}
 \end{eqnarray}
 \begin{eqnarray}
  \imath \kxzero (\dgammaw \yw - \czero) \tuwa + \tvwa \dgammaw & = & - \dfrac{\imath \kxzero \barT_w \tpwa}{\gamma \Ma^2} + \barT_w \func{\barT_w} \frac{d^2 \tuwa}{d \ywsq},
  \label{eq:eq49}
 \end{eqnarray}
 \begin{eqnarray}
  0 & = & - \frac{d \tpwa}{d \yw},
  \label{eq:eq410}
 \end{eqnarray}
  \begin{eqnarray}
  \imath \kxzero (\dgammaw \yw - \czero) \twwa & = & - \dfrac{\imath \kzzero \barT_w \tpwa}{\gamma \Ma^2} + \barT_w \func{\barT_w} \frac{d^2 \tww}{d \ywsq},
  \label{eq:eq49a}
 \end{eqnarray}
 Here, $\dgammaw$ 
 is the strain rate 
 at the wall, $y=0$. In equation \ref{eq:eq49} and \ref{eq:eq49a}, 
 the viscosity and thermal conductivity have been approximated by their values at the wall.
 The error due to these 
 approximations is $O(\Re^{-\frac{1}{3}})$. As is usual in boundary layer approximations, the cross-stream momentum equation, 
 \ref{eq:eq410}, reduces to the condition that the pressure gradient is zero. The mass and momentum equations are not 
 dependent on the temperature perturbation $\tTw$, and therefore these can be solved independently. It is convenient to
 add $\kxzero \times$ equation \ref{eq:eq49} and $\kzzero \times$ equation \ref{eq:eq49a}, to obtain,
 \begin{eqnarray} \lefteqn{
  \imath \kxzero (\dgammaw \yw - \czero) (\kxzero \tuwa + \kzzero \twwa) + \kxzero \tvwa \dgammaw} & & \nonumber \\
  & = & - \dfrac{\imath ( \kxzero^2 + \kzzero^2) \barT_w \tpwa}{\gamma \Ma^2} + \barT_w \func{\barT_w}\frac{d^2 (\kxzero \tuwa + \kzzero \twwa)}{d \ywsq}.
 \end{eqnarray}
 The $\yw$ derivative of the above equation is simplified using equation \ref{eq:eq48} and \ref{eq:eq410}, to obtain,
 \begin{eqnarray}
  \imath \kxzero (\dgammaw \yw - \czero) \frac{d (\kxzero \tuwa + \kzzero \twwa)}{d \yw} & = & \barT_w \func{\barT_w} \frac{d^3 (\kxzero \tuwa +
  \kzzero \twwa)}{d \ywcu}.
  \label{eq:eq412}
 \end{eqnarray}
 Here, we have neglected the derivatives of the mean quantities with respect to the cross-stream distance,
 because the characteristic length for the variation
 of the mean quantities, which is the channel width, is $O(\Re^{\frac{1}{3}})$ larger than the length scale for the perturbations
 in the wall layer. The solution of equation \ref{eq:eq412} for $\kxzero \tuwa + \kzzero \twwa$ which decreases to zero for $\yw \rightarrow \infty$ is
a generalised Airy function of the first kind \citep[see][Appendix A2]{ref-drazin_reid-98},
 \begin{eqnarray}
 \kxzero \tuwa + \kzzero \twwa & = & C \mbox{Ai}\left(\left(\frac{\imath \dgammaw \kxzero}{\barT_w \func{\barT_w}} \right)^{\frac{1}{3}} 
  \left( \yw - \frac{\czero}{\dgammaw} \right), 1 \right), \label{eq:eq413} 
  \end{eqnarray}
  \begin{eqnarray}
    \tvwa & = & - \imath C \left(\frac{\barT_w^{\zeta + 1}}{\imath \dgammaw \kxzero} \right)^{\frac{1}{3}}  \mbox{Ai}\left(\left(\frac{\imath \dgammaw \kxzero}{\barT_w \func{\barT_w}} \right)^{\frac{1}{3}} 
  \left( \yw - \frac{\czero}{\dgammaw} \right), 2 \right). \label{eq:eq413a}
 \end{eqnarray}
 The constant $C$ in equation \ref{eq:eq413} is determined from the no-slip condition at the wall, $\tuzero + \tuw = 0$
 and $\twzero + \tww = 0$ at $y = 0$ using equations \ref{eq:eq47b} and equation \ref{eq:eq413},
 \begin{eqnarray}
  C & = & - \imath \left( \mbox{Ai}\left(\mbox{} - \left(\frac{\imath \kxzero}{\dgammaw^2 \barT_w \func{\barT_w}} \right)^{\frac{1}{3}} \czero, 1 \right) \right)^{-1}. \label{eq:eq413b}
 \end{eqnarray}

 After the solution \ref{eq:eq413a} for $\tvw$ at the wall is obtained, $\czero$ is determined from equation
 \ref{eq:rayleighfiniteb} using the adjoint method and the value of $\tvbone$ at the wall. The adjoint operator $\cLfadj$ (equation \ref{eq:operatorfinite}) can be obtained as, 
 \begin{eqnarray}
  \lefteqn{\int_{y_l}^{y_h} dy \tvzeroadj \cLf(\tvbone)} & & \nonumber \\
  & = & \left. \frac{\baru}{\chizero} \left( \tvzeroadj \frac{d \tvbone}{d y} 
  - \tvone \frac{d \tvzeroadj}{d y} \right) \right|_{y_l}^{y_h} - \left. \frac{\tvzeroadj \tvbone}{\chizero} \frac{d \baru}{d y} 
  \right|_{y_l}^{y_h} + \int_{y_l}^{y_h} dy \tvbone \cLfadj(\tvzeroadj)
 \end{eqnarray}
 where the adjoint operator $\cLfadj$ is,
 \begin{eqnarray}
  \cLfadj(\tvzeroadj) & = & \frac{d}{d y} \left( \frac{\baru}{\chizero} \frac{d \tvzeroadj}{d y} \right) +
  \frac{1}{\chizero} \frac{d \baru}{d y} \frac{d \tvzeroadj}{d y} - \frac{(\kxzero^2 + \kzzero^2) \baru \tvzeroadj}{\barT}. \label{eq:eq48a}
 \end{eqnarray}
It is easily verified that the solution of the adjoint equation, \( \cLadj(\tvzeroadj) = 0 \) can be expressed in terms
of the solution \( \tvbzero \) of the homogeneous equation \( \cL(\tvbzero) = 0 \),
\begin{eqnarray}
 \tvzeroadj & = & \frac{\tvbzero}{\baru}. \label{eq:eq501}
\end{eqnarray}
 Multiplying equation \ref{eq:rayleighfiniteb} by $\tvzeroadj$ and integrating over the width of the channel, we obtain,
 \begin{eqnarray} \lefteqn{
  \left. \frac{\baru}{\chizero} \left( \tvzeroadj \frac{d \tvbone}{d y} 
  - \tvbone \frac{d \tvzeroadj}{d y} \right) \right|_{y_l}^{y_h} - \left. \frac{\tvzeroadj \tvbone}{\chizero} \frac{d \baru}{d y} 
  \right|_{y_l}^{y_h}} & & \nonumber \\
  & = & \int_{y_l}^{y_h} d y \, \tvzeroadj (\czero \cIcf(\tvbzero) + \kxone \cIxf(\tvbzero) + \kzone \cIzf(\tvbzero)),
\label{eq:eq502}
 \end{eqnarray}
 where \( \cIcf(\tvbzero), \cIxf(\tvbzero) \) and \( \cIzf(\tvbzero) \) are given in equations
 \ref{eq:inhomfinitec}-\ref{eq:inhomfinitez}.
 After substituting equation \ref{eq:eq501} for \( \tvzeroadj \) into \ref{eq:eq502}, and using the condition that \( \baru = 0 \) and \( \tvbzero= 0 \) at the wall, equation \ref{eq:eq502} is simplified as,
 \begin{eqnarray}
  \left. - \frac{\tvbone}{\chizero} \frac{d \tvbzero}{d y} \right|_{y_l}^{y_h} 
  & = & \int_{y_l}^{y_h} d y \, \tvzeroadj (\czero \cIcf(\tvbzero) + \kxone \cIxf(\tvbzero) + \kzone \cIzf(\tvbzero)).
\label{eq:eq503a}
 \end{eqnarray}

 The first correction to the velocity \( \tvbone \) in the second term on the right in
 equation \ref{eq:eq502} is determined from the first correction to the no-penetration conditions at the walls, \( \tvbone +\tvw = 0 \).
 \begin{eqnarray}
  \tvbone & = & \mbox{} - \tvwa = \mbox{} \left(\frac{\barT_w^{\zeta + 1}}{\imath \dgammaw \kxzero} \right)^{\frac{1}{3}} \frac{\mbox{Ai}\left(\mbox{} - \left(\dfrac{\imath \kxzero}{\dgammaw^2 \barT_w \func{\barT_w}} \right)^{\frac{1}{3}} \czero, 2 \right)}{
   \mbox{Ai}\left(\mbox{} -  \left( \dfrac{\imath \kxzero}{\dgammaw^2 \barT_w \func{\barT_w}} \right)^{\frac{1}{3}} \czero, 1 \right)}. \label{eq:eq504}
 \end{eqnarray}
Equation \ref{eq:eq503a}, with the substitutions \ref{eq:eq501} and \ref{eq:eq504}, is an implicit equation for \( \czero \), which has to be solved numerically in order to determine \( \czero \) as a function
of \( \kxone \) and \( \kzone \).

\begin{figure}
\begin{center}
  \subfigure[$k_x - k_{x0}$ vs. $c_R$]{{\label{fig-disp-match-m2-cr}}\includegraphics[width=0.65\textwidth]{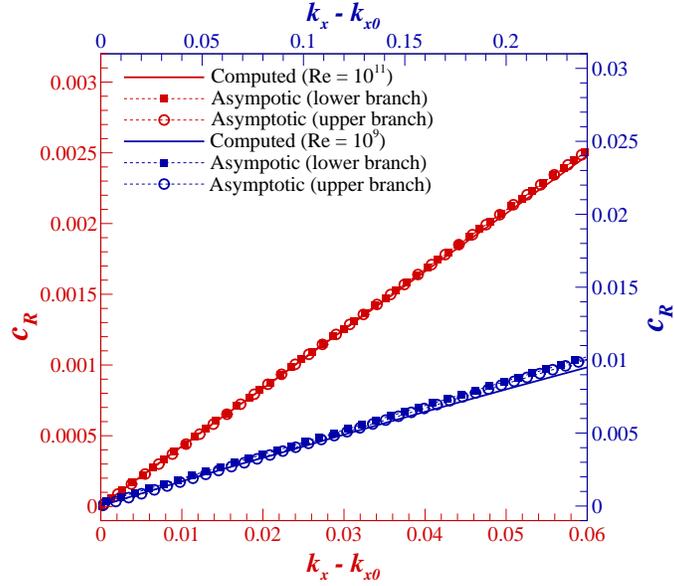}}
  \subfigure[$k_x - k_{x0}$ vs. $c_I$]{{\label{fig-disp-match-m2-ci}}\includegraphics[width=0.65\textwidth]{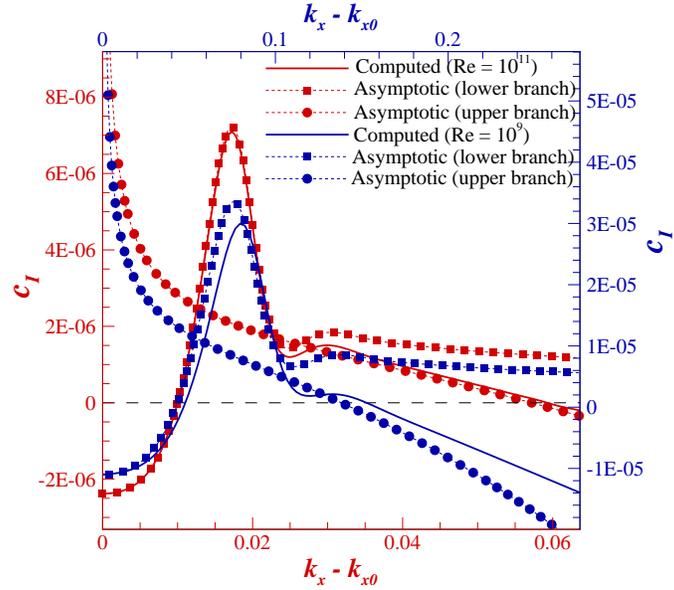}}
  \caption{Wave speed calculated from the asymptotic analysis for lower and upper branches compared with the numerical result of the linear stability equation at \( \Ma = 2 \) for the finite wave number mode II.} 
  \label{fig-disp-match-m2}
  \end{center}
\end{figure}

The results of the asymptotic analysis are compared with numerical results in figure \ref{fig-disp-match-m2} for mode II. The numerical solution is obtained by solving the full viscous stability equations for two-dimensional disturbances. Here, 
\( \kxzero \) is the stream-wise wave number at which \(c_R = 0 \) for the two-dimensional disturbance. Figure 
\ref{fig-disp-match-m2} (a) shows that the real part of the growth rate is in quantitative agreement with the numerical
solution. This is not surprising, because the real part of the growth rate is the solution of the leading order
inviscid equation \ref{eq:rayleighfinitea}, which is the same for the lower and upper branches. It was shown, in figure \ref{fig-invisc-drp-mode-2-3-comp} in section \ref{sec:channelflow}, that the inviscid calculations accurately predict the variation of the real part
of the growth rate with \( \kx \). Figure \ref{fig-disp-match-m2} (b) shows that the variation of the \( c_I \) is also
quantitatively predicted by the lower branch asymptotic analysis in the low wave number region, and especially in the vicinity 
of the lower branch, which is the wave number at which \( c_I \) crosses zero from below. There is some discrepancy between
the lower branch asymptotic results and the numerical results at higher wave number. This is because as $k_x$ becomes larger, the scalings for the lower branch asymptotic solution are no longer valid, and the scalings shift to
those for the upper branch asymptotic solution. However, figure \ref{fig-disp-match-m2}
shows that the numerical results are accurately captured by the asymptotic analysis in the wave number range of relevance
to the lower branch.

 \subsection{Lower branch small wave number}
 \label{subsec:lowerbranchsmall}
 For the lower branch of the small wave number modes, the relations in table \ref{tab:tablescal} are
 solved to obtain \( \alpha_c = \alpha_u = \alpha_w = \frac{2}{7}, \alpha_k = \frac{1}{7} \).
  The expressions for the velocity and pressure are provided in table \ref{tab:table1}.
The scaling of the pressure, density and temperature perturbations in the wall layer  is determined using the same procedure as that for the finite wave number modes, discussed after table \ref{tab:table1}. Here, the density and temperature perturbations are \( O(1) \), and the pressure perturbation is \( 
O(\Re^{- \frac{2}{7}}) \). Therefore, the density and temperature perturbations are large compared
to the pressure perturbation, and the leading order density and temperature perturbations are related
as \( \barrho \tT + \barT \trho = 0 \).
 
 The solution of the leading order Rayleigh equation \ref{eq:rayleighsmalla}, subject to the boundary conditions \( \tvbzero = 0 \) at the walls, is
 \begin{eqnarray}
  \tvbzero & = & \frac{\baru}{\baru_w'}, \label{eq:eq47aa} \\
  \kxzero \tubzero + \kzzero \twbzero & = & \frac{\imath}{\baru_w'} \frac{d \baru}{d y}, \label{eq:eq47bb}
 \end{eqnarray}
 where \( \baru_w' \) is the velocity gradient at the wall. In equation \ref{eq:eq47aa}, the normalisation condition \( (d \tvbzero/d y) = 1 \) is used. This is the same as
 that used in the solution \ref{eq:eq47a} for finite wave number modes.
 
 The growth rate \( \czero \) is determined from equation \ref{eq:rayleighsmallb} using the solvability condition.
 The adjoint of the linear operator \( \cLs \) (equation \ref{eq:operatorsmall}) is defined as,
 \begin{eqnarray}
  \cLsadj(\tvzeroadj) & = & \frac{d}{d y} \left( \frac{\baru}{\chizero} \frac{d \tvzeroadj}{d y} \right) +
  \frac{1}{\chizero} \frac{d \baru}{d y} \frac{d \tvzeroadj}{d y}. \label{eq:eq48b}
 \end{eqnarray}
 The solution of the homogeneous equation \( \cLsadj(\tvzeroadj) = 0 \) is, 
\begin{eqnarray}
 \tvzeroadj & = & \frac{\tvzero}{\baru} = \frac{1}{\baru_w'}. \label{eq:eq48c}
\end{eqnarray}
The equivalent of equation \ref{eq:eq503a} for the small wave number modes, after substituting
equations \ref{eq:eq47aa} and \ref{eq:eq48c}, is
 \begin{eqnarray}
  \left. - \frac{\tvbone}{\chizero} \frac{d \baru}{d y} \right|_{y_l}^{y_h} 
  & = & \int_{y_l}^{y_h} d y \, (\czero \cIcs(\tvbzero) + \cIks(\tvzero) (\kxzero^2 + \kzzero^2)),
\label{eq:eq503b1}
 \end{eqnarray}

\begin{figure}
\begin{center}
  \subfigure[$k_x$ vs. $c_R$]{{\label{fig-disp-match-m1-cr}}\includegraphics[width=0.65\textwidth]{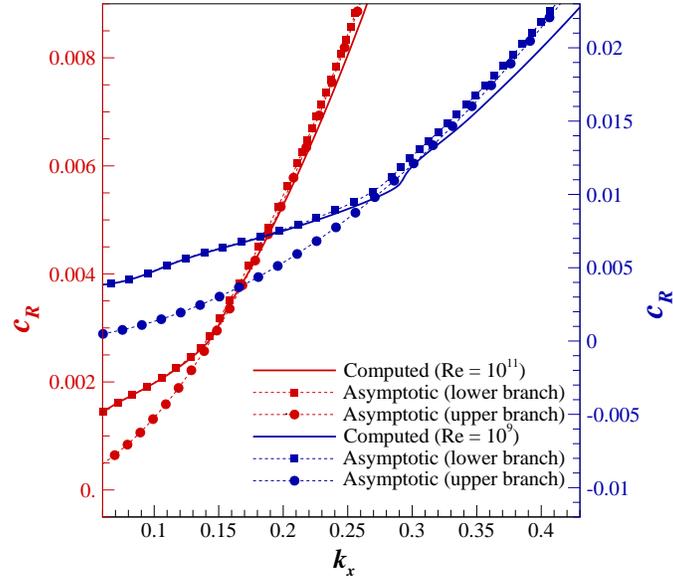}}
  \subfigure[$k_x$ vs. $c_I$]{{\label{fig-disp-match-m1-ci}}\includegraphics[width=0.65\textwidth]{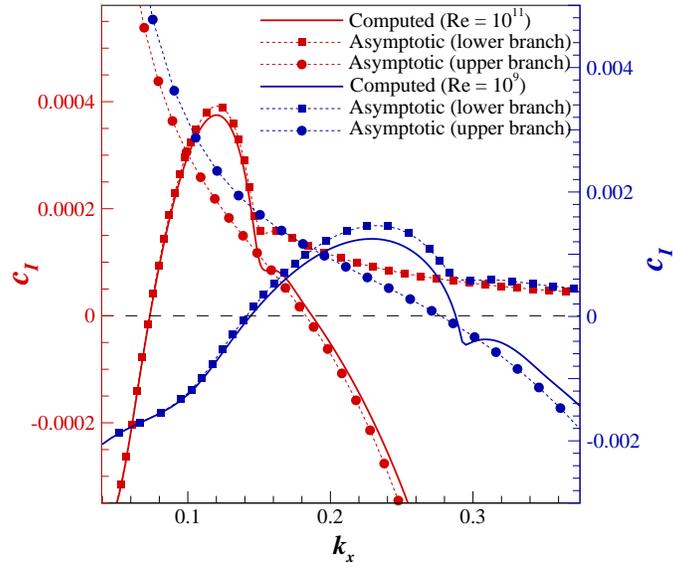}}
  \caption{Wave speed calculated from the asymptotic analysis for lower and upper branches compared with the numerical result of the linear stability equation at \(\Ma = 2 \) for the small wave number T-S mode.}
  \label{fig-disp-match-m1}
\end{center}
\end{figure}

 The equations for the wall layer are identical to equations \ref{eq:eq48}-\ref{eq:eq49a} when
 expressed in the scaled form. Consequently, the solution for \( \tvwa \) is given by equations
 \ref{eq:eq413a} and \ref{eq:eq413b}. The first correction to the no-penetration condition, 
 \( \tvbone = - \tvw \), is substituted into equation \ref{eq:eq503b1} to determine the growth rate 
 \( \czero \). The solution for the real and imaginary parts of the wave speed for \( \Re = 10^9, 10^{11}
 \) and \( \Ma = 2 \) for the small wave number modes are shown in figure \ref{fig-disp-match-m1}.
 Not surprisingly, the asymptotic solution for the real part of the wave number is in excellent agreement 
 with the numerical solution. For the imaginary part of the wave speed, the asymptotic result
 is in good agreement at low wave number, and the transition wave number is quantitatively captured
 by the asymptotic analysis.

 \subsection{Upper branch finite wave number}
 \label{subsec:upperbranchfinite}

  For the upper branch finite wave number modes, the relations in table \ref{tab:tablescal}
  are solved to obtain \( \alpha_u = \alpha_c = \alpha_k = \frac{1}{5}, \alpha_w = \frac{2}{5} \).
  The wave speed is \( O(\Re^{- \frac{1}{5}}) \) smaller than the flow velocity, and the viscous
  boundary layer thickness is \( O(\Re^{-\frac{2}{5}}) \) smaller than the channel width.
 The expressions for the velocity, density and pressure fields are summarised in table
 \ref{tab:table2}. Similar to the case of the lower branch analysis, the pressure perturbation
 is much smaller than the density and temperature perturbations here as well.
 \begin{table}
 \begin{center}
 \begin{tabular}{|l|c|c|} \hline
  & Finite wave number {mode} & Small wave number {mode}\\ \hline
  & $\alpha_u = \alpha_c = \alpha_k = \frac{1}{5}, \alpha_w = \frac{2}{5}$ & 
  $\alpha_u = \alpha_c = \frac{2}{11}, \alpha_k = \frac{1}{11}, \alpha_w = \frac{4}{11}$ \\ \hline
  $\kx$ &  $\kxzero + \Re^{-\frac{1}{5}} \kxone$ & $\Re^{-\frac{1}{11}} \kxzero$ \\ 
    $\kz$ & $\kzzero + \Re^{-\frac{1}{5}} \kzone$ & $\Re^{-\frac{1}{11}} \kzzero$ \\ 
    $c$ & $\Re^{-\frac{1}{5}} \czero + \Re^{-\frac{2}{5}} \cone$ & $\Re^{- \frac{2}{11}} \czero
    + \Re^{-\frac{4}{11}} \cone$ \\ \hline
    \multicolumn{3}{|c|}{Bulk flow} \\ \hline
$\tu$ & $\tubzero + \Re^{-\frac{1}{5}} \tubone$ & $\tubzero + \Re^{- \frac{2}{11}} \tubone$ \\ 
$\tv$ & $\tvbzero + \Re^{- \frac{1}{5}} \tvbone + \Re^{- \frac{2}{5}} \tvbtwo$ & $\Re^{-\frac{1}{11}} \tvbzero + \Re^{- \frac{3}{11}} \tvbone + \Re^{-\frac{5}{11}} \tvbtwo$ \\ 
$\tw$ & $\twbzero + \Re^{-\frac{1}{5}} \twbone$ & $\twbzero + \Re^{- \frac{2}{11}} \twbone$ \\ \hline
\multicolumn{3}{|c|}{Intermediate layer} \\ \hline
$y$ & $\Re^{-\frac{1}{5}} \yi$ & $\Re^{- \frac{2}{11}} \yi$ \\ 
$\baru$ & $\Re^{-\frac{1}{5}} \baru_w' \yi + \frac{1}{2} \Re^{-\frac{2}{5}} \baru_w'' \yisq$ 
& $\Re^{-\frac{2}{11}} \baru_w' \yi + \frac{1}{2} \Re^{-\frac{4}{11}} \baru_w'' \yisq$ \\ 
$\tu$ & $\tuizero + \Re^{-\frac{1}{5}} \tuione$ & $\tuizero + \Re^{-\frac{2}{11}} \tuione$ \\ 
$\tv$ & $\Re^{-\frac{1}{5}} \tvizero + \Re^{-\frac{1}{5}} \tvione$ &
$\Re^{-\frac{3}{11}} \tvizero + \Re^{-\frac{5}{11}} \tvione$ \\ 
$\tw$ & $\tvizero + \Re^{-\frac{1}{5}} \tvione$ & 
$\twizero + \Re^{-\frac{2}{11}} \twione$ \\ 
$\tp$ & $\Re^{-\frac{1}{5}} \tpizero + \Re^{-\frac{2}{5}} \tpione$ 
& $\Re^{-\frac{2}{11}} \tpizero + \Re^{-\frac{4}{11}} \tpione$ \\ \hline
\multicolumn{3}{|c|}{Viscous wall layer} \\ \hline
$y$ & $\Re^{-\frac{2}{5}} \yw$ & $\Re^{- \frac{4}{11}} \yw$ \\ 
$\baru$ & $\Re^{- \frac{2}{5}} \baru_w' \yw$ & $\Re^{- \frac{4}{11}} \baru_w' \yw$ \\ 
$\tu$ & $\tuw$ & $\tuw$ \\ 
$\tv$ & $\Re^{- \frac{2}{5}} \tvw$ & $\Re^{-\frac{5}{11}} \tvw$ \\ 
$\tw$ & $\tww$ & $\tww$ \\ 
$\tp$ & $\Re^{-\frac{1}{5}} \tpw$ & $\Re^{-\frac{2}{11}} \tpw$ \\ 
$\tT$ & $\tTw$ & $\tTw$ \\ 
$\trho$ & $\trhow$ & $\trhow$ \\ \hline
 \end{tabular}
\caption{\label{tab:table2} Scalings for the cross-stream co-ordinate, velocity, density, temperature and pressure fields
for the upper branch asymptotic analysis.}
 \end{center}
\end{table}

The equation for \( \tvbzero \) is identical to equation \ref{eq:rayleighfinitea}, and the boundary conditions are \( \tvbzero = 0 \) at the walls. This provides  the same solution for \( \kxzero \) 
and \( \kzzero \) as that for the lower branch. 
The solutions for \( (\tubzero, \tvbzero, \twbzero) \) are the same as \ref{eq:eq47a} and 
\ref{eq:eq47b} for the lower branch. 

As explained in section \ref{subsubsec:nopenetration}, the second correction \( \tvbtwo \) for the bulk flow is
comparable to the velocity in the wall layer \( \tvw \). The first correction \( \tvbone \)
is much larger than the velocity in the wall layer, and so the equation \ref{eq:rayleighfiniteb}
is solved subject to the condition that \( \tvbone = 0 \) at the wall. The leading order wave
speed \( \czero \) is determined from the solvability condition equation \ref{eq:eq503a} with \( \tvbone = 0 \),
\begin{eqnarray}
   \czero & = & - \frac{\int_{y_l}^{y_h} d y \, \tvzeroadj (\kxone \cIxf(\tvbzero) + \kzone \cIzf(\tvbzero))}{
   \int_{y_l}^{y_h} d y \, \tvzeroadj \cIcf(\tvbzero)}, \label{eq:eq55a}
\end{eqnarray}
where the adjoint \( \tvzeroadj \) is defined in equation \ref{eq:eq501}, and \( \cIcf, \cIxf \)
and \( \cIzf \) are defined in equations \ref{eq:inhomfinitec}-\ref{eq:inhomfinitez}.
Note that \( \czero \) is real, because \( \tvbzero, \cIcf(\tvbzero), \cIxf(\tvbzero) \) and \(
\cIzf(\tvbzero) \) are all real.
Therefore, the stability is determined by the imaginary part of the first correction to the wave
speed, \( \cone \). 

Though \( \czero \) has been relatively easily obtained from equation \ref{eq:eq55a},  
\( \cone \) has to be determined by solving equation \ref{eq:rayleighfiniteb}. 
If \( \czero \) is positive, equation \ref{eq:rayleighfiniteb} for \( \tvone \) contains
a singularity at the critical point \( y = y_c \) where \( \baru = \Re^{-\frac{1}{5}} \czero \), that is, the 
flow velocity is equal to the wave speed. In order to resolve this singularity, it is necessary to consider 
viscous effects within a critical layer of thickness \( \Re^{-\frac{1}{3}} \) around the critical point. 
To incorporate the effect of viscosity around the critical point, we consider an intermediate layer of thickness \( \Re^{- \frac{1}{5}} \) which incorporates the location where \( \baru = \Re^{-
\frac{1}{5}} \czero \).
The mean velocity and the perturbations to the pressure and velocity are expressed as
shown in table \ref{tab:table2}. Here, the stream-wise and span-wise velocity perturbations 
are \( O(1) \), comparable to those in the bulk. The cross-stream velocity perturbation
is \( O(\Re^{-\frac{1}{5}}) \) to satisfy the mass conservation condition. From the 
stream-wise momentum conservation equation \ref{u_mode}, shown in appendix \ref{secapp:normalmodeeqs}, the largest contribution to the 
pressure is \( O(\Re^{-\frac{1}{5}}) \).

The scaled co-ordinate is defined as \( \yi = \Re^{-\frac{1}{5}} y \) in the intermediate layer
within which the wave speed is equal to the flow velocity. Instead of using the mass conservation
equation \ref{rho_mode}, it is preferable to use the pressure form of the combined mass and energy 
conservation equations, \ref{eq:eq214a} (in appendix \ref{sec:inviscidflow} and \ref{secapp:normalmodeeqs}). In this equation, the first term on the left scales as
\( \Re^{-\frac{2}{5}} \), because the pressure is \( O(\Re^{- \frac{1}{5}}) \) and the mean velocity
is \( O(\Re^{-\frac{1}{5}}) \). The term on the right is \( O(\Re^{- \frac{3}{5}}) \), because
the characteristic length is \( O(\Re^{-\frac{1}{5}}) \). Therefore, the pressure term on the 
left and the terms on the right do not appear in either the \( O(1) \) or the \( O(\Re^{- \frac{1}{5}})
\) mass conservation equations,
\begin{eqnarray}
 \imath \kxzero \tuizero + \frac{d \tvizero}{d \yi} + \imath \kzzero \twizero & = & 0, \label{eq:eq611} \\
 \frac{d \tvione}{d \yi} + \imath \kxzero \tuione + \imath \kxone \tuizero +  
 \imath \kzzero \twione + \imath \kzone \twizero & = & 0. \label{eq:eq612}
 \end{eqnarray}
In the stream-wise momentum conservation equation \ref{u_mode}, the viscous term on the right 
is \( O(\Re^{-\frac{3}{5}}) \). Therefore, the viscous terms do not 
appear in the leading order or first correction to the equations, which are,
 \begin{eqnarray}
 \barrhow \imath \kxzero (\dgammaw \yi - \czero) \tuizero + \barrhow \dgammaw \tvizero & = & - \dfrac{\imath \kxzero}{\gamma \Ma^2} \tpizero, \label{eq:eq613}
 \end{eqnarray}
 \begin{eqnarray} 
 \imath (\dgammaw \yi - \czero)(\barrhow (\kxzero \tuione + \kxone \tuizero) + \barrhowp \yi \kxzero \tuizero) 
 & + & \barrhow\imath \kxzero (\tfrac{1}{2} \ddgammaw \yisq - \cone) \tuizero \nonumber \\  
 \mbox{} + \yi (\barrhow \ddgammaw + \barrhowp \dgammaw) \tvizero + \barrhow \dgammaw \tvione 
& = & - \dfrac{\imath (\kxzero \tpione + \kxone \tpizero)}{\gamma \Ma^2}.
 \label{eq:eq614}
\end{eqnarray}
In the cross-stream momentum conservation equation \ref{v_mode}, the inertial term on the 
left \( \imath \kx (\baru - c) \tv \sim \Re^{- \frac{2}{5}} \), because both the mean velocity
and the cross stream velocity perturbation are \( O(\Re^{-\frac{1}{5}}) \). The pressure gradient
\( (d \tp/d y) \) is \( O(1) \), because \( \tp \sim \Re^{- \frac{1}{5}} \) and \( y \sim
\Re^{- \frac{1}{5}} \). Therefore, the inertial terms are not present in the \( O(1) \) or
\( O(\Re^{- \frac{1}{5}}) \) cross-stream momentum equations,
\begin{eqnarray}
0 & = & - \frac{d \tpizero}{d \yi}, \label{eq:eq615} \\
0 & = & - \frac{d \tpione}{d \yi}. \label{eq:eq616} 
\end{eqnarray}
The scalings in the span-wise momentum equations are similar to those in the stream-wise
momentum equations \ref{eq:eq613} and \ref{eq:eq614},
\begin{eqnarray}
 \barrhow \imath \kxzero (\dgammaw \yi - \czero) \twizero & = & - \dfrac{\imath \kzzero}{\gamma \Ma^2} \tpizero, \label{eq:eq617} \\
\imath (\dgammaw \yi - \czero)[\barrhow (\kxzero \twione + \kxone \twizero) 
 & & \nonumber \\  
+ \barrhowp \yi \kxzero \twizero] + \barrhow\imath \kxzero (\tfrac{1}{2} \ddgammaw \yisq - \cone) \twizero
 & = & - \dfrac{\imath (\kzzero \tpione + \kzone \tpizero)}{\gamma \Ma^2}. \label{eq:eq618}
\end{eqnarray}

For the leading order solution, \( \kxzero \times \) equation \ref{eq:eq613} and \( \kzzero \times \) equation
\ref{eq:eq617} are added to obtain,
\begin{eqnarray}
 \barrhow \imath \kxzero (\dgammaw \yi - \czero) ( \kxzero \tuizero + \kzzero \twizero) + \barrhow
 \kxzero \dgammaw \tvizero & = & - \dfrac{\imath
 (\kxzero^2 + \kzzero^2) \tpizero}{\gamma \Ma^2} \tpizero, \nonumber \\ & & \label{eq:eq619}
\end{eqnarray}
Differentiating equation \ref{eq:eq619} with respect to \( \yi \), and using the mass conservation
equation \ref{eq:eq611} and equation \ref{eq:eq615} for the derivative of the pressure, we obtain,
\begin{eqnarray}
 \barrhow \imath \kxzero (\dgammaw \yi - \czero) \frac{d (\kxzero \tuizero + \kzzero \twizero)}{d \yi} & = & 0.
\label{eq:eq620} 
 \end{eqnarray}
The solution of the above equation that satisfies the condition \( \tvizero = 0 \) at
\( \yi = 0 \) and matches with the bulk solution for \( \tvbzero \), 
equation \ref{eq:eq47b}, is
\begin{eqnarray}
 \kxzero \tuizero + \kzzero \twizero & = & \imath, \label{eq:eq621} \\
 \tvizero & = & \yi, \label{eq:eq622} \\
 \tpizero & = & \frac{\imath \gamma \Ma^2 \czero \kxzero \barrhow}{\kxzero^2 + \kzzero^2}, \label{eq:eq623} \\
\tuizero & = & \frac{\imath}{\kxzero} + \frac{\imath \kzzero^2 \czero}{
 \kxzero (\kxzero^2 + \kzzero^2)(\dgammaw \yi - \czero)}, \label{eq:eq624} \\
 \twizero & = & \mbox{} - \frac{\imath \kzzero \czero}{(\kxzero^2 + \kzzero^2)(\dgammaw \yi - \czero)}.
 \label{eq:eq625}
\end{eqnarray}
It should be noted that there is no singularity in the solutions \ref{eq:eq621} and \ref{eq:eq622}, though
the solutions for \( \tuizero \) and \( \twizero \) have pole singularities at the location where the 
flow velocity is equal to the wave speed; these are to be regularised by including viscous effects. The singularity 
is not present for purely two dimensional disturbances without a span-wise component.

For the first correction to the solution in the intermediate layer, \( \kxzero \times \) equation \ref{eq:eq614} and \( \kzzero \times \) equation \ref{eq:eq618} are added, and the resulting equation is differentiated with respect to \( \yi \), to obtain,
\begin{eqnarray}
 \barrhow \imath \kxzero (\dgammaw \yi - \czero) \frac{d (\kxzero \tuione + \kzzero \twione)}{d \yi}
 + \barrhowp \kxzero \czero + \barrhow \kxzero \ddgammaw \yi 
  & & \nonumber \\ \mbox{} + \frac{\barrhow \dgammaw \czero \kzzero (\kxone \kzzero - \kzone \kxzero)}{ (\kxzero^2+\kzzero^2)(\dgammaw \yi - \czero)}
& = & 0. \label{eq:eq626}
\end{eqnarray}
Here, the derivatives of the pressure are set to zero in accordance with equations
\ref{eq:eq615} and \ref{eq:eq616},
we have substituted equation \ref{eq:eq621} for \( (\kxzero \tuizero + \kzzero \twizero) \), 
\ref{eq:eq622} for \( \tvizero \), and equation \ref{eq:eq624} for \( \tuizero \) and
\ref{eq:eq625} for \( \twizero \). Equation \ref{eq:eq626} is solved to obtain,
\begin{eqnarray}
 \kxzero \tuione + \kzzero \twione
 & = & \imath \left(\frac{\ddgammaw \yi}{\dgammaw} +
 \frac{\kzzero \czero (\kxzero \kzone - \kzzero \kxone)}{\kxzero (\kxzerosq
 + \kzzerosq)} \left( \frac{1}{\dgammaw \yi - \czero} + \frac{1}{\czero} \right) \right. 
 \nonumber \\ & & \mbox{} + \left. \left(
 \frac{\barrhow' \czero}{\barrhow \dgammaw} + 
 \frac{\ddgammaw \czero}{\dgammaw^2} \right) \log{(\xi/(- \czero))}
 \right), \label{eq:eq628}
\end{eqnarray}
where \( \xi = (\dgammaw \yi - \czero) \). Here, the constant of integration is set so that the 
no-slip condition is satisfied at the surface \( \yi = 0 \). From the mass conservation equation
\ref{eq:eq612}, the equation for \( \tvione \) is,
\begin{eqnarray}
 \frac{d \tvione}{d \yi} & = & \frac{\ddgammaw \yi}{\dgammaw}
 + \frac{\kxone \kxzero + \kzone \kzzero}{\kxzerosq + \kzzerosq}
+ \left(
 \frac{\barrhow' \czero}{\barrhow \dgammaw} + 
 \frac{\ddgammaw \czero}{\dgammaw^2} \right) \log{(\xi/(- \czero))}
 \label{eq:eq629}
\end{eqnarray}
In the solution, we substitute \( \log{(\yi - (\czero/\dgammaw))} = \log{|\yi - (\czero/\dgammaw)|}\) 
for \( \yi - (\czero/\dgammaw) > 0 \) and \( \log{(\yi - (\czero/\dgammaw))} = \log{|\yi - (\czero/\dgammaw)|} -\imath \pi\) for \( \yi - (\czero/\dgammaw) < 0$. Due to this phase shift in the complex plane, \( \tvione \) is complex in general.
The logarithmic singularity in the tangential velocity solution \ref{eq:eq628} at $\zeta = 0$ requires a viscous correction in the form of an internal critical layer within the intermediate layer. However, since the normal velocity solution is still regular at the leading and first correction order, the viscous corrections around the critical point are not necessary for the purpose of calculating the leading order wave speed and hence have not been calculated in this analysis. 

The first correction to the growth rate \( \cone \) has to be determined from an integral condition similar
to equation \ref{eq:eq503a}. The quantity of interest in these equations is \( \mbox{Im}(\cone) \),
the imaginary part of \( \cone \) which determines the stability. For this, it is sufficient to consider the 
equation for the imaginary part of the second correction to the velocity, \( \mbox{Im}(\tvbtwo) \). In the equivalent
of equations \ref{eq:eq503a} for \( \tvbtwo \), all inhomogeneous terms are real unless
they contain \( \cone \) or \( \tvbone \), both of which have imaginary parts. Therefore, the equation for the imaginary part of \( \tvbtwo \) is,
\begin{eqnarray}
 \cLf(\mbox{Im}(\tvbtwo)) & = & \czero \cIcf(\mbox{Im}(\tvbone)) + \kxone \cIxf(\mbox{Im}(\tvbone)) 
 + \kzone \cIzf(\mbox{Im}(\tvbone)) \nonumber \\ & &  + \mbox{Im}(\cone) \cIcf(\tvbzero),
 \label{eq:eq61}
\end{eqnarray}
where \( \mbox{Im}() \) is the imaginary part, and the functions \( \cIcf, \cIxf \) and 
\( \cIzf \) are given in equations \ref{eq:inhomfinitec}-\ref{eq:inhomfinitez}. 
The equation \ref{eq:eq61} is multiplied by 
\( \tvzeroadj \) (equation \ref{eq:eq501}), and integrated across the width of the channel to obtain
an equation similar to \ref{eq:eq503a} after some simplification and using the condition 
\( \tvbzero = 0 \) at the wall,
 \begin{eqnarray}
  \left. \mbox{} - \frac{\mbox{Im}(\tvbtwo)}{\chizero} \frac{d \tvbzero}{d y} \right|_{y_l}^{y_h} 
  & = & \int_{y_l}^{y_h} d y \, \tvzeroadj [\czero \cIcf(\mbox{Im}(\tvbone)) + \kxone \cIxf(\mbox{Im}(\tvbone)) 
  \nonumber \\ & & \mbox{} + \kzone \cIzf(\mbox{Im}(\tvbone)) + \mbox{Im}(\cone) \cIcf(\tvbzero)].
\label{eq:eq62}
 \end{eqnarray}
In the above equation, all terms on the right are known with the exception of \( 
\mbox{Im}(\cone) \). There are two types of terms in the integrals of \( \cIcf, \cIxf \)
and \( \cIzf \) in equation \ref{eq:eq62}, the terms that are exact differentials
(the first terms on the right in equations \ref{eq:inhomfinitec}-\ref{eq:inhomfinitez}) 
and those that are not exact differentials (the second terms on the right in equations \ref{eq:inhomfinitec}-\ref{eq:inhomfinitez}). The simplification of these terms is
illustrated for the first term on the right side of equation \ref{eq:eq62}.
\begin{eqnarray} \lefteqn{
 \int_{y_l}^{y_h} d y \, \tvzeroadj \cIcf(\mbox{Im}(\tvbone))} & & \nonumber \\ & = &
 \int_{y_l}^{y_h} d y \, \left[ \tvzeroadj \frac{d}{d y} \left( \cA(\mbox{Im}(\tvbone)) +
 \frac{1}{\chi_0} \frac{d (\mbox{Im}(\tvbone))}{d y} \right) - \frac{(\kxzerosq+\kzzerosq)
 \tvzeroadj \mbox{Im}(\tvbone)}{\barT} \right]\nonumber \\
 & = &  \mbox{} - \int_{y_l}^{y_h} d y \left[ \frac{d \tvzeroadj}{d y} \left(\cA(\mbox{Im}(\tvbone)) + 
 \frac{1}{\chizero} \frac{d(\mbox{Im}(\tvbone))}{d y} \right)
 + \frac{(\kxzerosq+\kzzerosq) \tvzeroadj \mbox{Im}(\tvbone)}{\barT} \right] \nonumber \\
 & & \mbox{} + \left. \tvzeroadj \cA(\mbox{Im}(\tvbone)) \right|_{y_l}^{y_h} +
 \left. \frac{\tvzeroadj}{\chizero} \frac{d (\mbox{Im}(\tvbone))}{d y} \right|_{y_l}^{y_h}.
 \label{eq:eq62aa}
\end{eqnarray}
In the integrals on the right side of equation \ref{eq:eq62aa}, the imaginary part
of \( \tvbone \) is non-zero only in a region of thickness \( O(\Re^{-\frac{1}{5}}) \)
at the wall, and the value of the integral is \( O(\Re^{-\frac{1}{5}}) \). The last 
two terms in equation
\ref{eq:eq62aa} are evaluated at the boundaries of the domain. In the first of these
terms, \( \tvzeroadj \) is finite at the walls, and the function \( \cA \) is proportional
to \( \baru^3 \) which decreases to zero at the walls. Therefore, the only non-zero
contribution is due to the last term on the right in equation \ref{eq:eq62aa}. In a similar
manner, it can be shown that the integrals of the terms proportional to \( \kxone \)
and \( \kzone \) in equation \ref{eq:eq62} are also zero, and the equation \ref{eq:eq62} 
reduces to,
 \begin{eqnarray}
  \left. \mbox{} - \frac{\mbox{Im}(\tvbtwo)}{\chizero} \frac{d \tvbzero}{d y} \right|_{y_l}^{y_h} 
  & = &   \left. \frac{\tvzeroadj}{\chizero} \frac{d (\mbox{Im}(\tvbone))}{d y} \right|_{y_l}^{y_h}
  + \int_{y_l}^{y_h} dy \tvzeroadj \mbox{Im}(\cone) \cIcf(\tvbzero).
\label{eq:eq62ab}
 \end{eqnarray}

The value of \( \tvbtwo \) at the wall is determined from the second correction to the zero velocity condition,
\begin{eqnarray}
\tvbtwo + \tvw & = & 0, \label{eq:eq62e}
\end{eqnarray}
at the wall. 

The equation for the velocity field in the wall layer is similar to equation \ref{eq:eq412}, but with one important
difference. Since the wall layer thickness is \( O(\Re^{-\frac{2}{5}}) \), while the wave speed is \( O(\Re^{- 
\frac{1}{5}}) \), the former can be neglected in comparison to the latter in the equivalent of equation
\ref{eq:eq412}. Thus, the analogue of equation \ref{eq:eq412} in the wall layer is,
\begin{eqnarray}
   - \imath \kxzero \czero \frac{d (\kx \tuw + \kz \tww)}{d \yw} & = & \barT_w \func{\barT_w} \frac{d^3 (\kx \tuw +
  \kz \tww)}{d \ywcu}.
  \label{eq:eq64}
 \end{eqnarray}
 This is easily solved to obtain,
\begin{eqnarray}
 (\kxzero \tuw + \kzzero \tww) & = & C_1 \exp{\left(- \sqrt{\dfrac{-\imath \kxzero \czero}{\barT_w \func{\barT_w}}} \yw\right)}, \nonumber \\
 \tvw & = & C_1 \sqrt{\frac{-\imath \barT_w \func{\barT_w}}{\kxzero\czero}}
 \exp{\left(- \sqrt{\dfrac{-\imath \kxzero \czero}{\barT_w \func{\barT_w}}} \yw\right)}. \label{eq:eq65}
\end{eqnarray}
The no-slip condition for the leading order tangential velocity, \( \kxzero (\tuw + \tuzero) + \kzzero (\tww + \twzero) = 0 \), provides 
the value \( C_1 = - \imath \). The boundary condition \ref{eq:eq62e} is then used to determine the imaginary part of 
\( \tvbtwo \) at the wall,
\begin{eqnarray}
 \tvbtwo & = &  -\sqrt{\frac{-\barT_w \func{\barT_w}}{\imath\kxzero\czero}} \label{eq:eq66}
\end{eqnarray}
The imaginary part of \( \tvtwo \) is substituted into equation \ref{eq:eq62} to determine the growth rate
\( \conei \).

The upper-branch asymptotic solution for real and imaginary parts of the wave speed are compared with numerical results
for high Reynolds number in figure \ref{fig-disp-match-m2}. The solution for the real part is in excellent 
agreement with the numerical results. In figure \ref{fig-disp-match-m2} (b), it is observed that the zero crossing of
the imaginary part of the wave speed is accurately captured by the asymptotic analysis. 
\subsection{Upper branch small wave number}
\label{subsec:upperbranchsmall}
For the small wave number modes on the upper branch, the relations in table \ref{tab:tablescal}
are solved to obtain,
   \begin{eqnarray}
    \alpha_c & = & \alpha_u = \tfrac{2}{11}, \: \: \alpha_k = \tfrac{1}{11}, \: \: \alpha_w = \tfrac{4}{11}. \label{eq:exp9}
   \end{eqnarray}
   The scaling of the velocity and pressure are provided in table \ref{tab:table2}.
   
   The equation for the leading order velocity \( \tvbzero \) is \ref{eq:rayleighsmalla},
   and the solution which satisfies the zero velocity boundary conditions is the same as
   that for the lower branch, equations \ref{eq:eq47aa} and \ref{eq:eq47bb}. The leading
   order wave speed is determined from the equation for the first correction, 
   \ref{eq:rayleighsmallb}. As noted
   in section \ref{subsec:upperbranchfinite}, the boundary condition for the first correction to the velocity 
   field is \( \tvbone = 0 \) for the upper branch. Therefore, the adjoint condition
   \ref{eq:eq503b1} can be used, with the condition that \( \tvbzero = 0 \) on the left
   side, to determine the first correction to the wave speed for the upper branch.
    \begin{eqnarray}
    \czero & = & \mbox{} - \frac{\int_{y_l}^{y_h} dy \cIks(\tvzero) (\kxzero^2 + \kzzero^2)}{\int_{y_l}^{y_h} d y \, \cIcs(\tvbzero)},
\label{eq:eq503b}
 \end{eqnarray}
 where \( \cIcs \) and \( \cIks \) are defined in equations \ref{eq:inhomsmallc} and 
 \ref{eq:inhomsmallk} respectively.
 
 The intermediate layer is of thickness \( \Re^{- \frac{2}{11}} \), and the velocity and 
 pressure fields in this intermediate layer are listed in table \ref{tab:table2}. The equations
 for the leading order and first correction to the velocity and pressure fields are obtained
 by substituting \( \kxone = \kzone = 0 \) in equations \ref{eq:eq611}-\ref{eq:eq618}. The solutions 
 for \( \tvbzero \) and \( (d \tvbone/d y) \) are obtained by substituting \( \kxone = \kzone = 0
 \) in equations \ref{eq:eq622} and \ref{eq:eq629}, respectively. The equation for the imaginary
 part of \( \tvbtwo \), analogous to equation \ref{eq:eq61} for finite wave number modes, is
 \begin{eqnarray}
 \cLs(\mbox{Im}(\tvbtwo)) & = & \czero \cIcs(\mbox{Im}(\tvbone)) + (\kxzerosq + \kzzerosq)
 \cIks(\mbox{Im}(\tvbone)) \nonumber \\ & & \mbox{} + \mbox{Im}(\cone) \cIcs(\tvbzero),
 \label{eq:eq61a}
\end{eqnarray}
where the operator \( \cLs \) is defined in equation \ref{eq:operatorsmall}, and the inhomogeneous
terms \( \cIcs \) and \( \cIks \) are defined in equations \ref{eq:inhomsmallc} and 
\ref{eq:inhomsmallk}. Equation \ref{eq:eq61a} is multiplied by the solution of the \( \tvzeroadj \)
(equation \ref{eq:eq48c}) of the adjoint operator \( \cLsadj \) (equation \ref{eq:eq48b}), and integrated across
the channel width, to obtain a relation similar to equation \ref{eq:eq62} for the finite wave number
modes, 
 \begin{eqnarray}
  \left. \mbox{} - \frac{\mbox{Im}(\tvbtwo)}{\chizero} \frac{d \baru}{d y} \right|_{y_l}^{y_h} 
  & = & \int_{y_l}^{y_h} d y \, [\czero \cIcs(\mbox{Im}(\tvbone)) + (\kxzero^2 +
  \kzzero^2) \cIks(\mbox{Im}(\tvbone)) \nonumber \\ & & \mbox{} + \mbox{Im}(\cone) \cIcs(\tvbzero)].
\label{eq:eq62b}
 \end{eqnarray}
 Here, we have substituted equation \ref{eq:eq622} for \( \tvbzero \) and \ref{eq:eq48c} for 
 \( \tvzeroadj \), respectively. 
 Equation \ref{eq:eq62b} is simplified
 using the same procedure as that for going from equation \ref{eq:eq62} to \ref{eq:eq62ab}.
 The equivalent of the integrals on the right in equation \ref{eq:eq62aa} are \( O(\Re^{-\frac{2}{11}}) \), since the imaginary part of \( \tvbone \) is non-zero in the region of
 thickness \( O(\Re^{-\frac{2}{11}}) \) from the wall. The equation \ref{eq:eq62b}
 is simplified as,
  \begin{eqnarray}
  \left. \mbox{} - \frac{\mbox{Im}(\tvbtwo)}{\chizero} \frac{d \tvbzero}{d y} \right|_{y_l}^{y_h} 
  & = &   \left. \frac{1}{\chizero} \frac{d (\mbox{Im}(\tvbone))}{d y} \right|_{y_l}^{y_h}
  + \int_{y_l}^{y_h} dy \mbox{Im}(\cone) \cIcs(\tvbzero).
\label{eq:eq62c}
 \end{eqnarray}
  The
 condition \ref{eq:eq66} is inserted into the left side of equation \ref{eq:eq62b},
 and equation \ref{eq:eq629}, with \( \kxone = \kzone = 0 \) is used for \( (d \tvbzero/dy) \) 
 at the boundary in order to determine the imaginary part of the first correction to the wave speed, \( \mbox{Im}(\cone) \).
 
 The results of the asymptotic calculation are compared with numerical results for \( \Re = 10^9, 10^{11} \)
 and \( \Ma = 2 \) in figure \ref{fig-disp-match-m1}. For the real part of the wave speed, there is
 a difference between the asymptotic and numerical results for small \( \kx \), but the two are 
 in agreement in the range of \( \kx \) where there is a transition from an unstable to a stable mode.
 For the upper branch, figure \ref{fig-disp-match-m1} also shows
 that the change in the sign of the imaginary part of the wave speed
 is also quantitatively predicted by the asymptotic analysis.
 
\begin{figure}
 \begin{center}
   \subfigure[]{\includegraphics[width=2.6in]{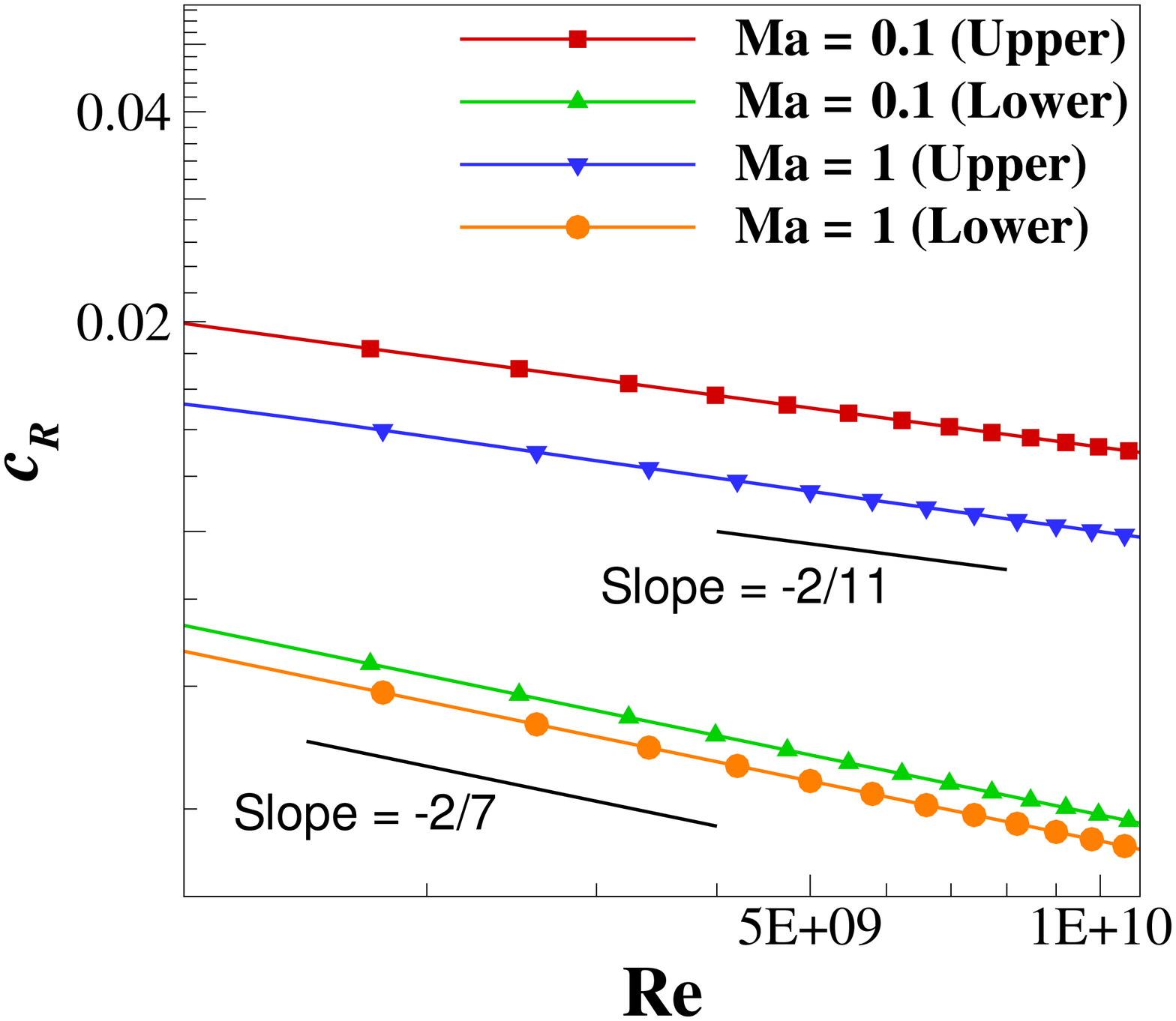}}
   \subfigure[]{\includegraphics[width=2.6in]{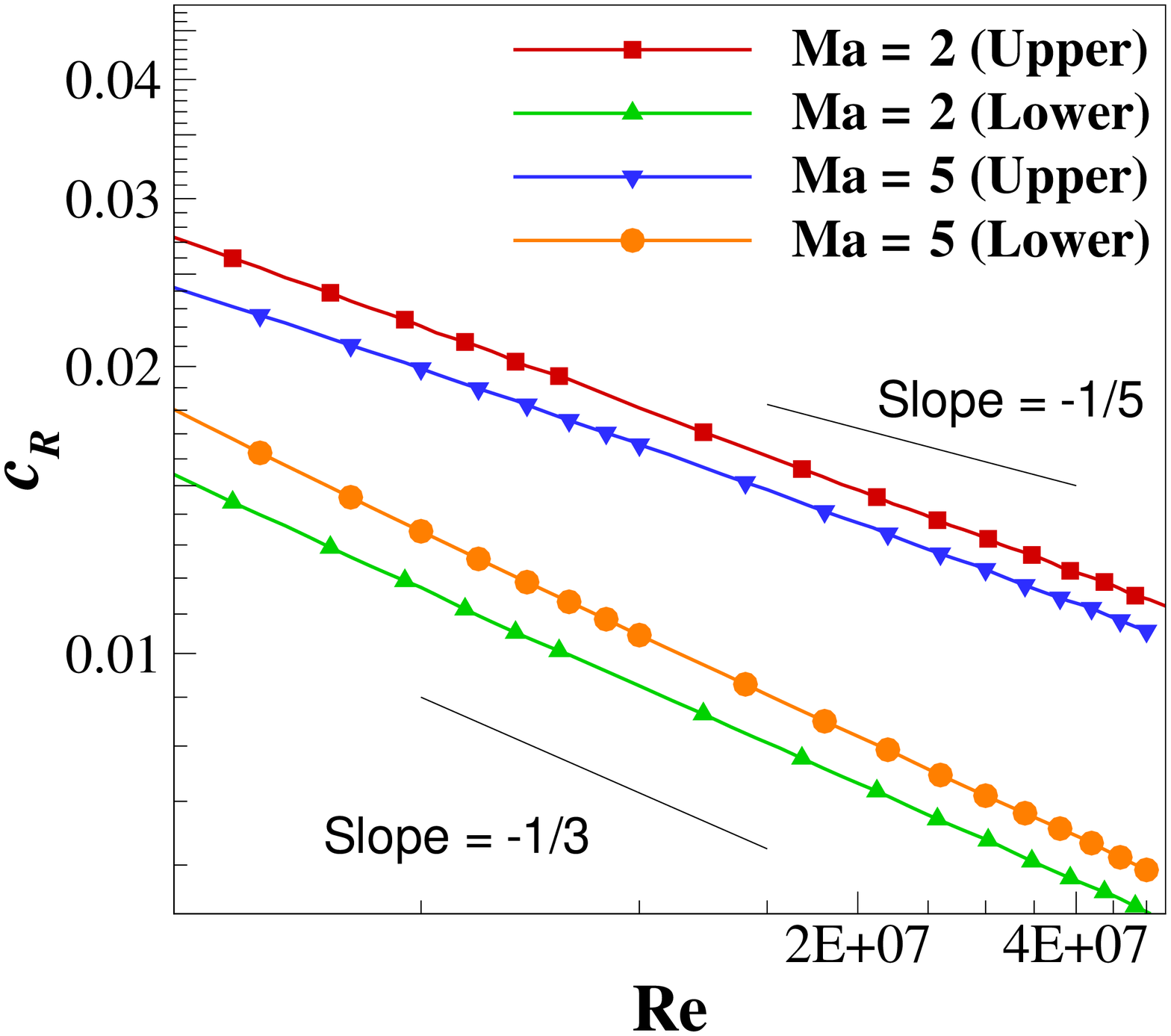}}
   \caption{\label{fig-neutral-visc}The real part of the wave speed $c_R$ as a function of 
  the Reynolds number (on a log-log scale) for the neutrally stable points ($c_I=0$) for T-S mode (a) and mode II (b).}
 \end{center}
\end{figure}

\begin{figure}
 \begin{center}
   \subfigure[]{\includegraphics[width=2.6in]{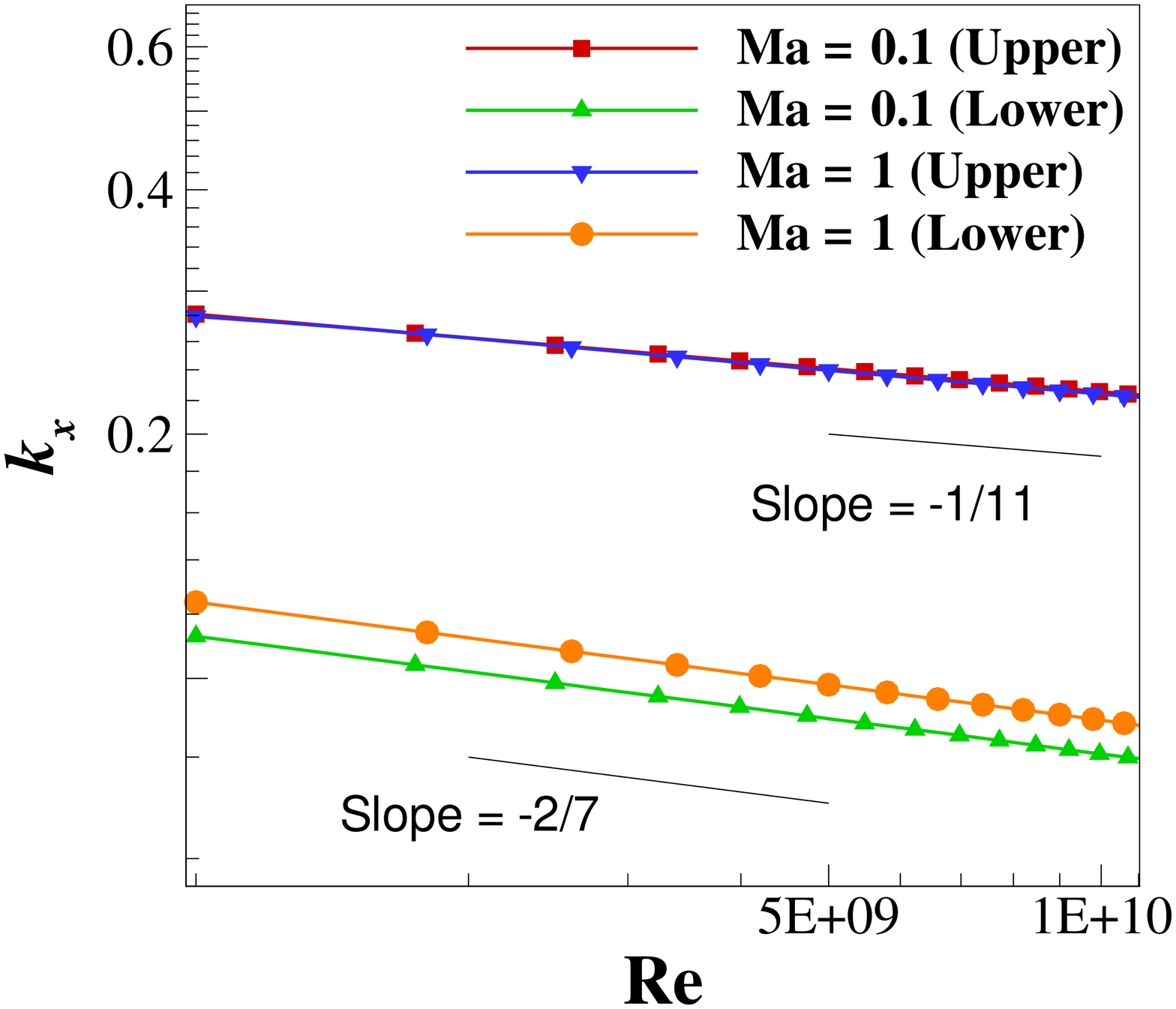}}
   \subfigure[]{\includegraphics[width=2.6in]{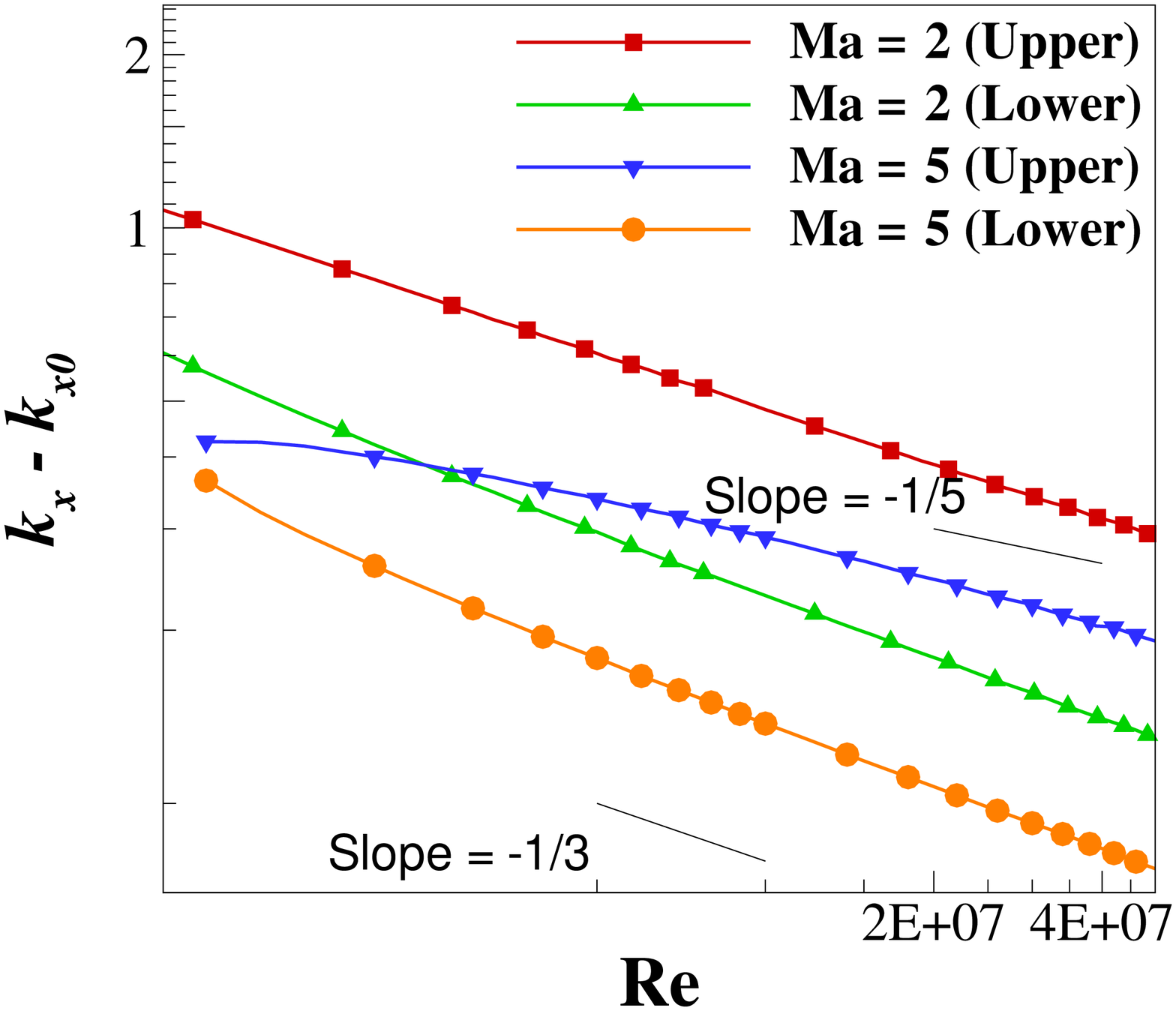}}
   \caption{\label{fig-neutral-visc-k}The wave-number as a function of 
  the Reynolds number (on a log-log scale) for the neutrally stable points ($c_I=0$) for T-S mode (a) and mode II (b).}
 \end{center}
\end{figure}

 The numerical results for the real
 part of the wave speed \( c_R \) for the neutrally stable modes are shown as a function of
 Reynolds number in figure \ref{fig-neutral-visc}. The wave number corresponding to the lower and upper branch points are also shown as a function of Reynolds number in figure \ref{fig-neutral-visc-k}. The scalings for the wave-number and wave  speed are in agreement with that predicted by the asymptotic analysis, and the difference in scaling for the lower and upper branches is clearly visible. It is also evident that the scaling of the finite wave number mode II is significantly different from the scaling for the small wave number T-S mode.

\section{Conclusions}
\label{sec:conclusion}

The present study shows that the instability of a compressible channel flow is qualitatively different
from that for an unbounded flow. The neutral modes have wave speed close to zero, and the destabilisation
is due to viscous effects in a critical layer close to the wall where the flow velocity is equal to the wave velocity. This is in contrast to the inviscid instability in a compressible boundary layer flow, where the wave speed of the neutral modes is finite, and flow is destabilised by an inviscid mechanism.
The unstable modes are of two types. The continuation of the Tollmien-Schlichting mode at finite Mach numbers, referred to as mode `0', is a small-wave number instability that has the lowest critical Reynolds number at low Mach numbers. The wave number for this instability decreases to zero in the limit of high Reynolds number. The second type is the higher modes present when the Mach number exceeds a threshold. These do not have a counterpart in the incompressible limit, and these are the most unstable modes at high Mach number. 
For the channel flow, these higher modes are categorised into a `lower' and an `upper' family, characterised by a monotonic increase and decrease in the real part of the wave  speed $\cR$, with streamwise wave-number $\kx$, respectively. At low wave-numbers, these modes appear as neutral waves in the inviscid limit with the wave speed outside the range of the minimum and maximum of the base velocity. The lower family, which causes the destabilisation of the flow, has a negative wave speed in the limit of low Reynolds number. As the Reynolds number is increased, at a specific wave-number, wave speed passes through zero and becomes positive. Since $\cR$ is  small but positive, a critical point appears close to the wall. The viscous effects in the critical layer destabilise the flow, even though the flow is stable in the inviscid approximation. 

A new Mach number criterion for the existence of these higher modes is derived in section \ref{sec:criticalmach}, using the condition that non-neutral modes can only exist is the real part of the wave-speed $\cR$ is within the interval $(\mbox{Min}(\baru),\mbox{Max}(\baru))$. The condition for the existence of neutral modes derived in section \ref{sec:conservationlaws} is, thereby, used to obtain a critical value of Mach number below which the higher modes are stable. This criterion can be generalised to any plane-parallel compressible shear flow in a bounded domain. It also applies to three-dimensional disturbances, and is used to obtain the critical wave-angles at a given Mach number (above the critical value for two-dimensional modes).

The asymptotic analysis for the lower and upper branches of the stability curve
was carried out in section \ref{sec:asymptoticanalysis}. The Reynolds number scaling
for the wave number and wave speed were derived on the basis of three criteria
explained in section \ref{subsec:scalingrelations}. These three criteria provide the 
complete set of relations listed in table \ref{tab:table1} that are used for 
determining all the exponents. These are different from the standard asymptotic
techniques, such as the WKBJ approximation, that are used to identify the upper
and lower branches.

By performing matched asymptotic analysis, we determine the lower and upper branch 
stability curves 
for general three-dimensional perturbations for both the Tollmien-Schlichting mode
and the compressible modes. 
In all cases, there is a critical layer within the 
flow where the wave speed is equal to the flow velocity, but the distance of
the critical layer from the wall decreases as a negative power of the Reynolds
number. For the lower branch, the wave speed is sufficiently small such that the critical 
point is within the viscous layer at the wall, whereas the critical layer
is well separated from the wall layer for the upper branch. 
The structure of the modes is similar to that
for two-dimensional perturbations, but there are important differences.
In the eigenfunctions for the upper branch in equations \ref{eq:eq621}-\ref{eq:eq625},
for example, the eigenfunctions \( \tuizero \) and \( \twizero \) are singular
at the critical point, though the sum \( \kxzero \tuizero + \kzzero \twizero \)
is finite. Despite these singularities, it was possible to identify a path to analytical
results for the leading order growth rates for the lower branch, and the leading
order and first correction to the growth rate for the upper branch. 

The present procedure recovers the exponents for the wave number for the
incompressible Tollmien-Schlichting mode, that is, 
\( k_x, k_z \propto \Re^{-\frac{1}{7}}, c \propto \Re^{- \frac{2}{7}} \) for the 
lower branch and \( k_x, k_z \propto \Re^{-\frac{1}{11}}, c \propto 
\Re^{-\frac{2}{11}} \) for the upper branch. The viscous wall layer thickness
scales as \( \Re^{-\frac{2}{7}} \) for the lower branch, while the viscous wall layer
and the inner critical layer thicknesses scale as \( \Re^{-\frac{4}{11}} \)
and \( \Re^{-\frac{2}{11}} \) respectively for the upper branch.
The scalings are different for the compressible finite wave number modes, which
are not present in an incompressible flow. 
The leading order wave speed scales as \( c \propto \Re^{-\frac{1}{3}} \) and 
\( c \propto \Re^{-\frac{1}{5}} \) for the lower and upper branches respectively.
The viscous wall layer for the lower branch scales as \( \Re^{-\frac{1}{3}} \),
while the viscous wall layer and the internal critical layer for the upper
branch scale as \( \Re^{-\frac{1}{2}} \) and \( \Re^{-\frac{1}{5}} \)
respectively. These were
shown to be in quantitative agreement with numerical simulations in the 
limit of high Reynolds number. This analysis provides two important 
extensions of the asymptotic analysis for the upper and lower branches,
namely the extensions to three dimensions and to compressible flows.

The analytical procedure for determining the lower and upper branches
can be easily extended to other types of compressible internal flows, since
the eigenfunctions of the most unstable modes are localised in a thin region
close to the wall. The criterion \ref{eq:macriterion} can be used to determine 
the minimum Mach number for a possible instability for any velocity and temperature
profile. The relation between the exponents in table \ref{tab:tablescal}, which
determine the Reynolds number scaling of the boundary layer thickness, velocity components 
and wave number, are general. The transition wave-number for the lower branch
is determined from the integral relation \ref{eq:eq503a}, which only requires the 
solution of the leading order inviscid flow velocity $\tvbzero$, the first correction
$\tvbone$ is calculated from the boundary condition \ref{eq:eq504} at the wall. 
For the upper branch, the integral condition \ref{eq:eq62ab} requires the values of $(d 
\tvbone/dy)$ and $\tvbtwo$ {at the boundaries}. The latter is determined from the {normal} velocity
boundary condition at the wall, \ref{eq:eq62e}, where the normal velocity in the wall layer is given
by \ref{eq:eq66}. The derivative $(d \tvbone/dy)$ is determined from the first correction to the continuity equation
for the inviscid outer flow, which requires the value of $(\imath \kxzero \tubone + \imath \kzzero \twbone)$; the latter is determined by matching with the intermediate layer solution for $(\imath \kxzero \tuione + \imath \kzzero \twione)$
given by equation \ref{eq:eq628}. The real part of this equation, which is due to the logarithmic
singularity in \ref{eq:eq628}, is sufficient for calculating the imaginary part of $(d \tvbone/dy)$,
which is then used to calculate imaginary part of $c_1^\ast$, which determines the stability. 
Therefore, it is only necessary to know the strain rate and temperature of the base flow at the wall to determine 
the stability of the system, and it is not necessary to compute the corrections to the inviscid velocity components in the bulk.

The authors thank the Science and Engineering Research Board, Department
of Science and Technology, Government of India, for financial support.

The authors report no conflict of interest.

\appendix   

\section{Inviscid stability}
\label{sec:inviscidflow}
To obtain the stability characteristics in the inviscid limit, the terms multiplied by $\Re^{-1}$ on the right side of equations
\ref{u_mode}-\ref{T_mode} (shown in appendix \ref{secapp:normalmodeeqs}) are neglected. The mass and energy balance equations, \ref{rho_mode} and \ref{T_mode}, can be combined by adding 
\( \barT \times \) equation \ref{rho_mode} and \( \barrho \times \) equation \ref{T_mode}, and 
dividing the resulting equation by \( \barrho \barT \), to obtain
one equation for the pressure perturbation in terms of the divergence of the velocity,
\begin{eqnarray}
\frac{\imath \kx (\baru - c) \tp}{\gamma \barrho \barT} + \left(\imath \kx \tu_x + \frac{d \tv}{d y} +
\imath k_z \tu_z \right) & = & 0. \label{eq:eq214a}
\end{eqnarray}
The above equation is used to express the divergence of the velocity in terms of the pressure
in the mass conservation equation \ref{rho_mode}, to 
obtain a relation between the density and pressure perturbations,
\begin{eqnarray}
 \trho & = & \frac{1}{\gamma} \frac{\tp}{\barT} + \frac{\barrho \tv}{\imath \kx (\baru - c) \barT} 
 \frac{d \barT}{d y}.
 \label{eq:eq214}
\end{eqnarray}
In equations \ref{eq:eq214a} and \ref{eq:eq214}, we have used the condition that the pressure in the cross-stream direction is invariant
in the base state, $(d \barp/d y) = 0$, and therefore $(d \barrho/d y) = - (\barrho/\barT)
(d \barT/dy)$ from the equation of state. In inviscid approximation for the mass conservation 
equation \ref{rho_mode}, equations \ref{eq:eq214}, \ref{v_mode} and \ref{w_mode} are used
to substitute for $\trho$, $\tu$ and $\tw$, respectively, to obtain,
\begin{eqnarray}
 \barrho \frac{d \tv}{d y} - \frac{\barrho \tv}{\baru - c} \frac{d \baru}{d y} + \frac{\imath k_x 
 \tp}{\gamma} \left( \frac{\baru - c}{\barT} - \frac{(\kx^2 + \kz^2)}{\kx^2 \Ma^2 (\baru - c)} \right)
 & = & 0. \label{eq:eq215}
\end{eqnarray}
Equation \ref{eq:eq215} can then be simplified to determine $\tp$ in terms of $\tv$,
\begin{eqnarray}
 \tp & = & \frac{\gamma \barrho \barT \Ma^2 \kx (\baru - c)}{\imath (\kx^2 + \kz^2) \chi} \left( \frac{d \tv}{d y} - 
 \frac{\tv}{\baru - c} \frac{d \baru}{d y} \right), \label{eq:eq216}
\end{eqnarray}
where
\begin{eqnarray}
 \chi & = & \barT - \frac{\Ma^2 \kx^2 (\baru - c)^2}{\kx^2 + \kz^2}. \label{eq:eq217}
\end{eqnarray}
This is substituted into the inviscid approximation for equation \ref{v_mode} to obtain the Rayleigh equation,
 \begin{eqnarray}
  \frac{d}{d y} \left( \frac{(\baru - c)}{\chi} \frac{d \tv}{d y} - \frac{\tv}{\chi} \frac{d \baru}{d y} \right) & = & \frac{(\kx^2 + \kz^2) (\baru - c) \tv}{\barT}. \label{eq:eq218}
 \end{eqnarray}


\subsection{General results}
\label{subsec:theorems}
Though it appears that equation \ref{eq:eq218} can be used to prove the Squire theorem,
this is misleading because the scaled mean velocity and temperature profiles depend on
the Mach number. If the growth rate is $c$ for three-dimensional perturbations with wave
numbers $(\kx, \kz)$ in the stream- and span-wise directions for Mach number $\Ma$, the 
growth rate is the same for two-dimensional perturbations with $\kaxsq = \kx^2 + \kz^2$
and for a lower Mach number $\Maasq = (\Ma^2 \kx^2/\kaxsq)$. However, the mean velocity
and temperature profiles change when the Mach number changes, and so it is not possible
to conclude that two-dimensional disturbances are always more unstable than three-dimensional
disturbances.

 When equation \ref{eq:eq218} is multiplied by $\tva$, the complex conjugate of $\tv$, divided by $\baru - c$ and simplified, we obtain,
 \begin{eqnarray}
 \tva\frac{d}{d y} \left( \frac{1}{\chi} \frac{d \tv}{d y} \right) - \frac{\tv \tva}{
 \baru - c} \frac{d}{d y}
 \left( \frac{1}{\chi} \frac{d \baru}{d y} \right) & = & \frac{(\kxsq + \kzsq) \tv \tva}{\barT}. 
\label{eq:eq12}
 \end{eqnarray}
 This is subtracted from its complex conjugate, and integrated over the width of the channel,
 to obtain
 \begin{eqnarray}
 \int_{y_l}^{y_h} d y \left[ \tva\frac{d}{d y} \left( \frac{1}{\chi} \frac{d \tv}{d y} \right) -
 \tv \frac{d}{d y} \left( \frac{1}{\chia} \frac{d \tva}{d y} \right)
 - \frac{\tv \tva}{\baru - c} \frac{d}{d y}
 \left( \frac{1}{\chi} \frac{d \baru}{d y} \right) 
+ \frac{\tv \tva}{
 \baru - \ca} \frac{d}{d y}
 \left( \frac{1}{\chia} \frac{d \baru}{d y} \right) \right]
 & = & 0 \nonumber \\ & & 
  \label{eq:eq13}
  \end{eqnarray}

Integrating the first term by parts twice, and using the zero normal velocity condition, the above relation can be simplified to,
\begin{equation}
\begin{aligned}
 \int_{y_l}^{y_h} dy \Big[ -4 \imath \Ma^2 c_I \tv &\frac{d}{dy}\left( \frac{(\baru - c_R)}{|\chi|^2} \frac{d\tva}{dy} \right) - \dfrac{4 \imath \Ma^2 (\baru - c_R) c_I|\tv|^2}{|\baru - c|^2} \dfrac{d}{dy} \left( \frac{(\baru - c_R)}{|\chi|^2} \frac{d\baru}{dy} \right) \\
	& + \dfrac{2 \imath c_I|\tv|^2}{|\baru - c|^2} \dfrac{d}{dy} \left( \frac{(\barT - \Ma^2 (\baru - c_R)^2 + \Ma^2 c_I^2)}{|\chi|^2} \frac{d\baru}{dy} \right)\Big] = 0,
\end{aligned}
\label{eq:gip_prop_1}
\end{equation}
where, $c_R$ and $c_I$ denote the real and imaginary parts of $c$ respectively, and the expression for $\chi$ is substituted from equation \ref{eq:eq217} for the simplification. To derive the equivalent of the Rayleigh theorem, we notice that for modes with $c_I \rightarrow 0$ in the inviscid limit, if $\mbox{Min}(\baru) < c_R < \mbox{Max}(\baru)$, in the region where $|\baru - c_R| \sim O(|c_I|)$, the first and the second term on the left hand side of equation \ref{eq:gip_prop_1} are $O(|c_I|^2)$ and $O(|c_I|)$ respectively but the last term becomes $ O(|c_I|^{-1})$ and remains unbalanced unless,
\begin{equation*} \frac{d}{dy} \left( \frac{1}{\barT} \frac{d \baru}{d y} \right) \sim O(|c_I|^2). \end{equation*}

This results in the following proposition,
 \begin{theorem}
An inviscid mode with \( c_I \rightarrow 0 \) and $\mbox{Min}(\baru) < c_R < \mbox{Max}(\baru)$ can exist in an flow only if
 \[
  \frac{d}{dy} \left( \frac{1}{\barT} \frac{d \baru}{d y} \right) \rightarrow 0
 \]
at the location where $\baru = c_R$.
\label{theorem:proposition1}
\end{theorem}
This is the equivalent of the Rayleigh inflection point theorem for an incompressible
flow often referred to in previous literature as the generalised inflection point (GIP) criteria \citep{ref-lees_lin-46,ref-duck-94}. This criteria, however, is much restrictive than its incompressible counterpart since it serves as a necessary condition for the existence of a non-neutral mode if only \( c_I \rightarrow 0 \) and $\mbox{Min}(\baru) < c_R < \mbox{Max}(\baru)$ for the mode.
 
Bounds on the wave speed can be derived from equation \ref{eq:eq12} without having to make
approximations beyond the inviscid approximation.
The function $\tg$ is defined as $\tv/(\baru - c)$, and equation \ref{eq:eq218} is expressed
in terms of $\tg$,
\begin{eqnarray}
 \frac{d}{d y} \left( \frac{(\baru - c)^2}{\chi} \frac{d \tg}{d y} \right) & = & \frac{(\kxsq + \kzsq) (\baru - c)^2 \tg}{\barT}
\end{eqnarray}
This equation is multiplied by the complex conjugate $\tga$ and integrated across the channel,
\begin{eqnarray}
 \left. \left( \frac{(\baru - c)^2 \tga}{\chi} \frac{d \tg}{d y} \right) \right|_{y_l}^{y_h} -
 \int_{y_l}^{y_h} dy \left( \frac{(\baru - c)^2}{\chi} \left| \frac{d \tg}{d y} \right|^2
 + \frac{(\kxsq + \kzsq) (\baru - c)^2 |\tg|^2}{\barT} \right) & = & 0. \nonumber \\ & & 
 \label{eq:eq15}
\end{eqnarray}
The first term on the left is zero at both boundaries due to the zero normal-velocity condition. The remaining two terms can be simplified as,
\begin{eqnarray}
 \int_{y_l}^{y_h} dy \left( \Psi - (\baru - c)^2 \Phi  \right) & = & 0.
 \label{eq:eq16}
\end{eqnarray}
where the functions
\begin{eqnarray}
 \Psi & = & \frac{\Masq \kxsq |\baru - c|^4}{(\kxsq + \kzsq) |\chi|^2} \left| \frac{d \tg}{d y} \right|^2, \: \:
 \Phi = \left(\frac{\barT}{|\chi|^2} \left| \frac{d \tg}{d y} \right|^2 + \frac{(\kxsq + \kzsq) |\tg|^2}{\barT} \right),
 \label{eq:eq16a}
\end{eqnarray}
are positive throughout the domain.

The imaginary part of equation \ref{eq:eq16} is,
\begin{eqnarray}
 2 c_I \int_{y_l}^{y_h} dy (\baru - c_R) \Phi & = & 0. \label{eq:eq17}
\end{eqnarray}
From this, we obtain,
\begin{theorem}
For non-neutral modes with \( c_I \neq 0 \) for an inviscid flow, the real part of the wave speed  $c_R$ is bounded by the minimum and 
maximum of $\baru$.
\label{theorem:proposition2}
\end{theorem}

Equation \ref{eq:eq16} is multiplied by $\ca$, and the imaginary part of the resulting equation is,
\begin{eqnarray}
 c_I \int_{y_l}^{y_h} dy \left( \Psi
 + (|c|^2 - \baru^2) \Phi \right) & = & 0. \nonumber \\ & & \label{eq:eq18}
 \end{eqnarray}
 From this, we obtain,
 \begin{theorem}
  For a non-neutral mode with \( c_I \neq 0 \) in an inviscid flow, 
  $|c|^2$ is bounded by the minimum and maximum of $\baru^2$.
  \label{theorem:proposition3}
 \end{theorem}
 
 From the real and imaginary parts of equation \ref{eq:eq16}, it can be inferred that
 \begin{eqnarray}
  \int_{y_l}^{y_h} dy \baru \Phi & = & c_R \int_{y_l}^{y_h} dy \Phi, \label{eq:eq19} \\
  \int_{y_l}^{y_h} dy \baru^2 \Phi & = & \int_{y_l}^{y_h} dy ((c_R^2 + c_I^2) \Phi + \Psi). \label{eq:eq110}
 \end{eqnarray}
Using the identity
\begin{eqnarray}
\int dy (\baru - \mbox{Min}(\baru))(\baru - \mbox{Max}(\baru)) \Phi & < & 0, \label{eq:eq111}
\end{eqnarray}
where $\mbox{Min}(\baru)$ and $\mbox{Max}(\baru)$ are the minimum and maximum values of the mean velocity.
Substituting from equations \ref{eq:eq19} and \ref{eq:eq110} for the integrals of the $\baru \Phi$ and
$\baru^2 \Phi$, we get,
\begin{eqnarray}
 \int_{y_l}^{y_h} dy [((c_R^2 + c_I^2) - c_R (\mbox{Min}(\baru) + \mbox{Max}(\baru)) + \mbox{Min}(\baru) \mbox{Max}(\baru)) \Phi + \Psi] & < 
 & 0. \nonumber \\ & & \label{eq:eq111}
\end{eqnarray}
 Since the function $\Psi$ is always positive, this implies that
\begin{eqnarray}
 \int_{y_l}^{y_h} dy [(c_R - \mbox{$\frac{1}{2}$} (\mbox{Max}(\baru) + \mbox{Min}(\baru)))^2 + c_I^2 - \mbox{$\frac{1}{4}$} (\mbox{Max}(\baru) - \mbox{Min}(\baru))^2] \Phi & <  & 0. \nonumber \\ & & \label{eq:eq112}
\end{eqnarray}
 Since the function $\Phi$ is always positive, the above inequality implies that,
 \begin{theorem}
 For a non-neutral mode with \( c_I \neq 0 \) in an inviscid flow,
 \begin{eqnarray}
  (c_R - \mbox{$\frac{1}{2}$} (\mbox{Max}(\baru) + \mbox{Min}(\baru)))^2 + c_I^2 & < & \mbox{$\frac{1}{4}$} (\mbox{Max}(\baru) - \mbox{Min}(\baru))^2,
  \label{eq:eq113}
 \end{eqnarray}
 where \( \mbox{Max}(\baru) \) and \( \mbox{Min}(\baru) \) are the minimum and maximum values of the mean velocity.
 \label{theorem:proposition4}
 \end{theorem} 

This is the equivalent of the Howard semi-circle theorem for compressible flows derived previously by \cite{ref-eckart-63}. It should be noted that propositions \ref{theorem:proposition2} - \ref{theorem:proposition4} apply only for non-neutral modes
 in an inviscid flow, since the assumption \( c_I \neq 0 \) has been made in equations \ref{eq:eq17}, \ref{eq:eq18} 
 and \ref{eq:eq19}. For neutral modes with $c_I = 0$, equation \ref{eq:eq15} can be rewritten as,
 \begin{eqnarray}
 \int_{y_l}^{y_h} dy (\baru - c_R)^2 \left( \frac{1}{\chi} \left| \frac{d \tg}{d y} \right|^2
 + \frac{(\kxsq + \kzsq) |\tg|^2}{\barT} \right)  & = & 0. \label{eq:eq114}
 \end{eqnarray}

For span-wise perturbations imposed to the mean flow, the density, velocity and temperature
are expressed as,
 \begin{eqnarray}
 \rho & = & \barrho + \trho \exp{(\imath \kz z + s t))}, \label{eq:span1} \\
 \bu & = & \baru {\bf e}_x + \tbu \exp{(\imath \kz z + s t)}, \label{eq:span2} \\
 T & = & \barT + \tT \exp{(\imath \kz z + s t)}, \label{eq:span3} \\
 p & = & \barp + \tp \exp{(\imath \kz z + s t)}. \label{eq:span4}
\end{eqnarray}
The above equations can be simplified to express the density and velocity in terms of the temperature,
\begin{eqnarray}
 \tv & = & - \frac{1}{\barrho \gamma \Ma^2 s} \frac{d \tp}{d y}, \: \: \tu = \frac{1}{\barrho \gamma
 \Ma^2 s^2} \frac{d \baru}{d y} \frac{d \tp}{d y}, \: \: \tw = - \frac{\imath \kz \tp}{\barrho \gamma
 \Ma^2 s}, \label{eq:span5} \\
 \tT & = & \frac{(\gamma-1) \tp}{\barrho \gamma} + \frac{1}{\barrho \gamma \Ma^2 s^2} \frac{d \barT}{d y} 
 \frac{d \tp}{d y}, \: \: \trho = \frac{\tp}{\gamma \barT} - \frac{1}{\gamma \barT \Ma^2 s^2} \frac{d \barT}{d y}
 \frac{d \tp}{d y}. \label{eq:span6}
\end{eqnarray}
These are inserted into the mass conservation equation \ref{rho_mode} to obtain one second order equation for 
the pressure,
\begin{eqnarray}
 \frac{d}{d y} \left( \barT \frac{d \tp}{d y} \right) - (\Ma^2 s^2 + \barT \kz^2) \tp & = & 0.
 \label{eq:span7}
\end{eqnarray}
The zero normal velocity condition at the boundaries implies that $(d \tp/d y) = 0$ at the boundaries. If
we multiply the left side by the complex conjugate of $\tp$ and integrate across the channel, it is easy to 
show that $s^2$ has to be negative, and its magnitude has to be greater than the minimum of $(\barT \kz^2/
\Ma^2)$, for the existence of solutions, hence the following result,
 \begin{theorem}
 A purely span-wise normal mode disturbance in the inviscid limit is always neutrally stable.
 \label{theorem:proposition6}
 \end{theorem}

\section{Normal mode equations}
\label{secapp:normalmodeeqs}

The linearised mass, $x$-momentum, $y$-momentum, $z$-momentum, and temperature equations for normal modes imposed on a compressible plane parallel flow are, 
\begin{equation}
\label{rho_mode}
 \imath \kx ( \baru - c ) \trho + \frac{d \barrho}{dy} \tv + \barrho \left( \imath \kx \tu + \frac{d \tv}{dy} + \imath \kz \tw \right) = 0,
	\end{equation}
	\begin{eqnarray}
	\lefteqn{
\barrho \left(\imath \kx ( \baru - c ) \tu + \frac{d\baru}{dy} \tv \right)} & & 
\nonumber \\ & = & -\frac{\imath \kx \tp}{\gamma \Ma^{2}} + \frac{1}{\Re} \left[ \func{\barT} \left( \dfrac{d^2}{dy^2} - \kxsq - \kzsq \right) \tu + 
\dfrac{d F}{dT}\Bigg|_{T={\barT}}\dfrac{d^2 \baru}{d y^2} \tT \right. \nonumber \\ & & \left. \mbox{} + 
\dfrac{d F}{dT}\Bigg|_{T={\barT}}\dfrac{d \barT}{d y} \left( \dfrac{d \tu}{d y} + \imath \kx \tv \right) 
 + \left( \dfrac{d F}{dT}\Bigg|_{T={\barT}}\frac{d \tT}{d y} + \dfrac{d^2 F}{dT^2}\Bigg|_{T={\barT}}\frac{d \barT}{
 d y} \tT \right) \dfrac{d\baru}{dy} \right. \nonumber \\ & & \left. 
 + \imath \kx (\mu_b + \mbox{$\frac{1}{3}$}) \func{\barT} \left( \imath \kx \tu + \frac{d\tv}{dy} + \imath \kz \tw\right) \right] + \dfrac{1}{\Fr} \trho, 	\label{u_mode}
 	\end{eqnarray}
	\begin{eqnarray} \lefteqn{
\barrho \imath \kx ( \baru - c ) \tv} & & \nonumber \\ & = & -\frac{1}{\gamma \Ma^{2}} \frac{d\tp}{dy} + \frac{1}{\Re} \left[ \func{\barT} 
\left( \dfrac{d^2}{dy^2} - \kxsq - \kzsq \right) \tv + \imath \kx \dfrac{d F}{dT}\Bigg|_{T={\barT}}\dfrac{d\baru}{dy} \tT  \right. \nonumber \\ & &\left. \mbox{} 
+ 2 \dfrac{d F}{dT}\Bigg|_{T={\barT}}\dfrac{d \barT}{d y} \dfrac{d\tv}{dy} 
+ (\mu_b -\tfrac{2}{3})
\dfrac{d F}{dT}\Bigg|_{T={\barT}}\dfrac{d \barT}{d y} \left( \imath \kx \tu + \frac{d\tv}{dy} + \imath \kz
\tw \right) \right. \nonumber \\ & & \left. \mbox{} + (\mu_b + 
\mbox{$\frac{1}{3}$}) \func{\barT} \frac{d}{d y} \left( \imath \kx \tu + \frac{d \tv}{dy} + \imath
\kz \tw \right) \right], 
    \label{v_mode}
	\end{eqnarray}
	\begin{eqnarray} \lefteqn{
\barrho \imath \kx ( \baru - c ) \tw} & & \nonumber \\
& = & -\frac{\imath \kz \tp}{\gamma \Ma^{2}} + 
\frac{1}{\Re} \left[ \func{\barT} 
\left( \dfrac{d^2}{dy^2} - \kxsq - \kzsq \right) \tw 
+ \dfrac{d F}{dT}\Bigg|_{T={\barT}}\dfrac{d \barT}{d y} \left( \dfrac{d\tw}{dy} + \imath \kz \tv \right) \right. 
\nonumber \\ & & \left. +  \imath \kz(\mu_b + 
\mbox{$\frac{1}{3}$}) \func{\barT} \left( \imath \kx \tu + \frac{d \tv}{dy} + \imath
\kz \tw \right) \right], 
	\label{w_mode}
	\end{eqnarray}

	\begin{eqnarray} \lefteqn{
\barrho \left(\imath \kx ( \baru - c ) \tT + \frac{d\barT}{dy} \tv \right)} & & \nonumber \\
& = & - (\gamma - 1) \barrho \barT 
\left( \imath \kx \tu + \frac{d\tv}{dy} + \imath \kz \tw\right) 
+ \frac{\gamma}{\Re \Pr} \left[ \func{\barT} \left( \dfrac{d^2}{dy^2} - \kxsq - \kzsq \right)  \tT 
\right. \nonumber \\ & & \left. \mbox{} 
+ 2 \dfrac{d F}{dT}\Bigg|_{T={\barT}} \dfrac{d \barT}{d y} \dfrac{d \tT}{d y} + \left(\dfrac{d^2 F}{dT^2}\Bigg|_{T={\barT}} \left( \dfrac{d\barT}{dy}\right)^2  + \dfrac{d F}{dT}\Bigg|_{T={\barT}} \dfrac{d^2 \bar{T}}{d y^2} \right) \tT \right] 
	\nonumber \\ & & \mbox{}
	+ \dfrac{\gamma (\gamma - 1) \Ma^2}{\Re} \left[ \dfrac{d F}{dT}\Bigg|_{T={\barT}} \left( \dfrac{d\bar{u}}{dy}\right)^2 \tT + 2 \func{\barT} \dfrac{d\baru}{dy} \left( \dfrac{d \tu}{d y} + \imath \kx \tv \right) \right].
		\label{T_mode}
	\end{eqnarray}
	The equation of state is,
\begin{equation}
\label{p_mode}
    \tp = \barrho \tT + \barT \trho.
\end{equation}
An equation for the pressure perturbation, useful in some cases, is derived by adding \( \barT \)
times the equation \ref{rho_mode} for the density and \ref{T_mode}, and using the equation of
state \ref{p_mode} to identify the pressure perturbation,
	\begin{eqnarray} \lefteqn{
\imath \kx ( \baru - c ) \tp + \gamma \barrho \barT \left(\imath \kx \tu + 
\frac{d \tv}{d y} + \imath \kz \tw \right)} & & \nonumber \\
& = &  \frac{\gamma}{\Re \Pr} \left[ \func{\barT} \left( \dfrac{d^2}{dy^2} - \kxsq - \kzsq \right)  \tT 
\right. \nonumber \\ & & \left. \mbox{} 
+ 2 \dfrac{d F}{dT}\Bigg|_{T={\barT}} \dfrac{d \barT}{d y} \dfrac{d \tT}{d y} + \left(\dfrac{d^2 F}{dT^2}\Bigg|_{T={\barT}}\left( \dfrac{d\barT}{dy}\right)^2  + \dfrac{d F}{dT}\Bigg|_{T={\barT}} \dfrac{d^2 \bar{T}}{d y^2} \right) \tT \right] 
	\nonumber \\ & & \mbox{}
	+ \dfrac{\gamma (\gamma - 1) \Ma^2}{\Re} \left[ \dfrac{d F}{dT}\Bigg|_{T={\barT}} \left( \dfrac{d\bar{u}}{dy}\right)^2 \tT + 2 \func{\barT} \dfrac{d\baru}{dy} \left( \dfrac{d \tu}{d y} + \imath \kx \tv \right) \right].
		\label{pp_mode}
	\end{eqnarray}
Here, we have substituted \( (d \barp / d y) = \barT (d \barrho/dy) + \barrho (d \barT/dy) = 0 \)
for the mean flow, in order to eliminate the terms proportional to \( \tv (d \barrho/d y) \)
and \( \barrho \tv (d \barT/dy) \) on the left sides of equation \ref{rho_mode} and \ref{T_mode},
respectively.

\bibliographystyle{jfm}
\bibliography{ref}

\end{document}